\documentclass{JHEP3} 
\usepackage{amsmath}
\usepackage{epsfig}
\usepackage{amssymb,amsfonts}

\def\en{\mathbb{N}}
\def\zi{\mathbb{Z}}
\def\d{{\partial}}
\def\ni{\noindent}
\def\slr{$SL(2,\mathbb{R})$}
\def\slc{$SL(2,\mathbb{R})/U(1)$}
\def\be{\begin{eqnarray}}
\def\ee{\end{eqnarray}}
\def\nn{\nonumber}
\def\nst{\widetilde{NS}}
\def\nstr{$\mathrm{\nst}$}
\def\bb{${\cal L}_{1}$}
\def\gn{${\cal L}_{2}$}
\def\sl{${\cal L}_{sl}^{\pm}$}
\def\ntl{${\cal L}_{N=2}^{\pm}$}

\def\TL{\hfil$\displaystyle{##}$}
\def\TR{$\displaystyle{{}##}$\hfil}

\def\lbldef#1#2{\expandafter\gdef\csname #1\endcsname {#2}}
\def\eqn#1#2{\lbldef{#1}{(\ref{#1})}%
\begin{equation} #2 \label{#1} \end{equation}}
\def\eqalign#1{\vcenter{\openup1\jot
    \halign{\strut\span\TL & \span\TR\cr #1 \cr
   }}}

\def\href#1#2{#2}

\title{
\boldmath D-branes in N=2 Liouville theory and its mirror
\unboldmath}
\author{Dan Isra\"el${}^1$, Ari Pakman${}^{2,3}$ and
Jan Troost${}^1$
\\  ${}^1$ Laboratoire de Physique Th\'eorique
de l'\'Ecole Normale Sup\'erieure\thanks{Unit{\'e} mixte  du
CNRS et de l'Ecole Normale Sup{\'e}rieure,
UMR 8549.}  \\ 24, Rue Lhomond,  75231
Paris Cedex  05, France\\
 \\
 $ {}^2$ Racah Institute of Physics, The Hebrew University \\
Jerusalem 91904, Israel \\
\\
 $ {}^3$ Erwin Schr\"odinger International Institute for Mathematical Physics \\
 Boltzmanngasse 9, A-1090 Vienna, Austria \\
\\
E-mail:  \email{israel@lpt.ens.fr, pakman@phys.huji.ac.il, troost@lpt.ens.fr }
\\
}

\abstract{
We study D-branes in the mirror pair $N=2$ Liouville/supersymmetric
$SL(2,\mathbb{R})/U(1)$ coset superconformal field theories.
After revisiting the duality between the two models, we  build D0, D1 and D2 branes, on the basis of the boundary state
construction for the $H_{3}^+$ conformal field theory. We also construct
D0-branes in an orbifold that rotates the angular direction of the cigar.
We show how the poles of correlators associated to localized states and bulk
interactions naturally decouple in the one-point functions of localized and
extended branes. We stress the role played in the analysis of D-brane spectra
by primaries in $SL(2,\mathbb{R})/U(1)$ which are descendents of the parent theory.}

\preprint{
LPTENS-04/27 \\
RI-05-04
\\hep-th/0405259}

\begin{document}


\section{Introduction}
\label{intro}
We have learned  that it is very useful to study
non-perturbative objects in string theory, especially when they are
related to an implementation of holography.
This study has proved instrumental both for understanding
gauge theory physics, and for getting to grips with aspects
of quantum gravity. In this paper, we will concentrate
on constructing boundary conformal field theories for theories with $N=2$
supersymmetry and central charge $c>3$. These boundary conformal field
theories are arguably
the most important missing ingredient in the construction of
D-branes in non-compact curved string backgrounds with bulk supersymmetry.

The models we will study in detail are the $N=2$ Liouville
theory \cite{Girardello:1990sh,Kutasov:1990ua},
and the $SL(2,\mathbb{R})/U(1)$ super-coset \cite{Kazama:1988qp} theory.
These two theories
are known to be dual \cite{FZZ,Giveon:1999px}(see also \cite{Hori:2001ax,Tong:2003ik}),
and  are mapped to each other by mirror
symmetry.

Constructing D-branes
in these superconformal field theories will not only increase our
knowledge of D-branes in non-minimal $N=2$ superconformal field theories (see
also \cite{Eguchi:2003ik,McGreevy:2003dn,Ahn:2003tt}),
but it will also enable us to study holographic dualities in closer detail.
There are two conjectured instances of holography
where these D-branes are expected to be relevant.
The $N=2$ Liouville/$SL(2,\mathbb{R})/U(1)$ background is
is  related to a conjectured calculable version
of holography \cite{Giveon:1999px} which involves a bulk closed superstring
background~\cite{Israel:2004ir} dual to a non-gravitational non-local Little
String Theory \cite{Seiberg:1997zk}.
Note for instance that D1-branes stretching between NS5-branes (which can be
interpreted as the W-bosons of the Little String Theory) can be
constructed using these boundary conformal field theories.

Another natural area where the D-branes built in this model are
relevant is the study of matrix models for two-dimensional
non-critical superstrings \cite{McGreevy:2003dn} and Type $0$
strings in a 2D black hole. For the latter case, it was
conjectured in~\cite{Giveon:2003wn}, following the ideas of
\cite{Douglas:2003up}, that the decoupled theory of $N \rightarrow
\infty$ $D0$ branes of ZZ type in $N=2$ Liouville, leads to a
version of the matrix model of \cite{Kazakov:2000pm} with the
matrix eigenvalues filling symmetrically both sides of the
inverted harmonic oscillator potential. This matrix model would be
dual to 2D Type $0A$ string theory in the supersymmetric 2D black
hole background.

Our paper is structured as follows. We first review in section
\ref{bulk} the bulk theories, and the duality between $N=2$
Liouville theory and the supersymmetric coset. In section
\ref{boundary} we go on to discuss how to relate $H_{3}^+$ (i.e.
Euclidean $AdS_3$) boundary conformal field theories to the $N=2$
theories under consideration. In the following sections, we then
explicitly construct and study  D0-, D1- and D2-branes. We pause
in section \ref{orbifold} to explain how to extend our result to
an angular orbifold of the cigar.
 In the appendices, we collect several important remarks.
One concerns the fact that the \slc\ super-coset characters are
equal to the characters of the $N=2,c>3$ Virasoro
 algebra. Another
treats the Fourier transformation of the one-point functions, while a third
appendix analyzes the $SL(2,\mathbb{R})$ symmetry of general $N=2$ $c>3$ conformal
field theories.


\section{The bulk}
\label{bulk} In this section we will study aspects of
 the bulk theory in which the D-branes studied in this paper will be embedded.
We give original points of view on some of the topics treated in the
literature.
We first discuss the
spectrum of the  supersymmetric $SL(2,\mathbb{R})/U(1)$ conformal field
theory for concreteness. We then
study the nature of the duality between the susy coset and $N=2$
Liouville (as well as its bosonic counterpart). We next review the
pole structure of the bulk correlators  and
comment upon the pole structure of the one-point functions.
We finish by suggesting that from the perspective of
recent developments in non-rational conformal field theories,
the duality of the two theories is a case of dynamics being completely determined by
chiral symmetries.

\subsection{The bulk spectrum of the supersymmetric coset}
We consider a bulk theory with $N=2$ supersymmetry, namely the
axial $SL(2,\mathbb{R})/U(1)$ super-coset conformal field theory
\cite{Kazama:1988qp}. We
review the algebraic details of this $N=2$ algebra in Appendix
\ref{characters}. The central charge of the conformal field theory
is \be c = 3 + \frac{6}{k}
\,, \label{cencha} \ee where k is the level of the parent (total) \slr\
current algebra. The primaries $\Phi^j_{m,\bar{m}}$ in the bulk, coming from
\slr\ primaries with spin $j$, can belong either to the NS or R sector
of the $N=2$ algebra.
They have conformal dimensions and $N=2$
R-charges \cite{Giveon:2003wn,Israel:2004ir}
\be
\Delta_{j,w,n} &=& -\frac{j(j-1)}{k} + \frac{m^2}{k} +\frac{({\delta^R_{\pm}})^2}{8} \qquad  \qquad Q_{j,w,n} = \frac{2m}{k} +\frac{{\delta^R_{\pm}}}{2} \,, \nn  \\
\bar{\Delta}_{j,w,n} &=& -\frac{j(j-1)}{k} + \frac{\bar{m}^2}{k} +
\frac{({\delta^R_{\pm}})^2}{8} \qquad \qquad \bar{Q}_{j,w,n} =
-\frac{2\bar{m}}{k}  -\frac{{\delta^R_{\pm}}}{2} \,,
\label{cosetnumbers} \ee where \be m = \frac{n+kw}{2} \qquad
\qquad \bar{m} = -\frac{n-kw}{2} \,, \ee with $n,w \in
\mathbb{Z}$. The constant $\delta^R_{\pm}$ is zero in the $NS$
sector and $\pm 1$ in the R sector. The Ramond primaries appear
always in pairs due to the double degeneracy of the Ramond vacuum.
The numbers $m,\bar{m}$ are the eigenvalues of the left and right
elliptic generators of the $SL(2,\mathbb{R})$ Lie algebra. Notice
that the conformal dimension $\Delta_{j,w,n}$ is invariant under
$j \rightarrow -j+1$. The two corresponding primaries are related
through \be
\Phi^{-j+1}_{m,\bar{m}} &=& R^{NS/R^{\pm}}(-j+1, m,\bar{m}) \ \Phi^j_{m,\bar{m}} \,, \nn \\
&=& \frac{1}{R^{NS/R^{\pm}}(j, m,\bar{m})} \ \Phi^j_{m,\bar{m}} \,,
\label{reflection}
\ee
where the reflection  coefficients
$R^{NS/R^{\pm}}(j,m,\bar{m})$ are given by \cite{Giveon:2003wn}
\be
 R^{NS}(j,m,\bar{m}) &=&  \nu^{1-2j}
\frac{\Gamma(-2j+1)\Gamma(1+\frac{1-2j}{k})} {\Gamma(2j-1)
\Gamma(1-\frac{1-2j}{k})}
\frac{\Gamma(j+m)\Gamma(j-\bar{m})}{\Gamma(-j+1+m)\Gamma(-j+1-\bar{m})} \,,  \label{tpf} \\
 R^{R^{\pm}}(j,m,\bar{m}) &=&  \nu^{1-2j} \frac{\Gamma(-2j+1)\Gamma(1+\frac{1-2j}{k})} {\Gamma(2j-1) \Gamma(1-\frac{1-2j}{k})}
\frac{\Gamma(j+m \pm \frac12)\Gamma(j-\bar{m}\mp\frac12)}{\Gamma(-j+1+m\pm\frac12)\Gamma(-j+1-\bar{m}\mp\frac12)} \,, \nn \\
\label{tpframond} \ee for primaries of the NS and R sectors,
respectively.

The two-dimensional geometry of the axial coset
theory is a cigar \cite{Elitzur:cb, Mandal:1991tz},
 or Euclidean 2D black hole \cite{Witten:1991yr,Giveon:1991sy} . It has an asymptotic radius equal to $\sqrt{k
\alpha'}$, and the quantum number $n$ is interpreted in the axial
$SL(2,\mathbb{R})/U(1)$ coset as the momentum around the compact circle at
infinity. The number $w$ denotes the winding number of the
closed string states~\cite{Dijkgraaf:1991ba}, but can be viewed also as
the spectral flow parameter of the affine
$SL(2,\mathbb{R})$ algebra~\cite{Henningson:1991jc,Maldacena:2000hw}.

The spectrum of the super-coset includes both continuous
representations with $j=\frac{1}{2}+iP$ (and $P \in \mathbb{R}^+_0$), and
discrete {\it lowest-weight} representations ${\cal D}_{j}^{+}$
\cite{Dijkgraaf:1991ba} with values of $j \in \mathbb{R}$
satisfying
\be
\frac12 < j < \frac{k+1}{2} \,,  \label{unitbound}
\ee and \be
j + r = m\,, \qquad && j + \bar{r} = \bar{m} \,,
\ee where $r,\bar{r}$
are integers (half-integers) for the Neveu-Schwarz (Ramond)
sector. The bound  for $j$ (\ref{unitbound}) is stricter than
the unitarity bound on the coset
representations~\cite{Dixon:1989cg} as well as than the bound corresponding to
normalizable operators~\cite{Maldacena:2000hw}.
 The improved bound has been shown to apply in all physical settings
\cite{Giveon:1999px,Maldacena:2000hw,Hanany:2002ev,Israel:2003ry,Eguchi:2004yi,Israel:2004ir}.
Although coset primaries for discrete representations appear for all values of
$r,\bar{r}$, only those with $r,\bar{r} \geq 0$ correspond to
coset primaries inherited from \slr\ primaries, and only for them
expressions (\ref{cosetnumbers}) hold. The coset primaries with
$r,\bar{r} < 0$ arise from {\it descendants} of \slr\ discrete {\it
lowest-weight} representations ${\cal D}_{j}^{+}$, but they can be interpreted as
{\it primaries} coming from discrete {\it highest-weight}
representations ${\cal D}_{\frac{k+2}{2} -j}^{-}$ with  spin $\frac{k+2}{2} -j$. We discuss the
details of the $r,\bar{r} < 0$ primaries in this section, in
section~\ref{locbra} and in Appendix~\ref{characters}.\footnote{Note that when
  we take into account all
primaries including those with $r,r'<0$, there is no need to consider  both ${\cal D}^{+}$
and ${\cal D}^{-}$
representations, the ${\cal D}^{+}$ representations  being enough to cover all the spectrum.}

The spectrum of primaries that we have reviewed above follows from
studying the representation of the affine \slr\ algebra and from
the analysis of the modular invariant partition function of the
model~\cite{Hanany:2002ev,Eguchi:2004yi,Israel:2004ir}. The reflection
coefficients (\ref{tpf})-(\ref{tpframond}) are related to the dynamics of the
theory, i.e. the correlation functions, which we consider now.

\subsection{Towards the duality with $N=2$ Liouville: a bosonic ancestor}
Before discussing the duality between the susy coset \slc\ and
$N=2$ Liouville, we will look at its bosonic counterpart.
The  correlators of the bosonic \slc\ theory can be  obtained by
free field computations in \slr. For this one can use the Wakimoto
free field representation of the algebra\footnote{We will take the
\slr\ level $k+2$,
 which is more convenient in order to move to the susy case later. We take $\alpha'=2$.}
\begin{eqnarray}
j^+ & = & \beta \nonumber \\
j^3 & = &-\beta \gamma - \frac{1}{Q} \partial \phi \nonumber \\
j^- & = & \beta \gamma^2 + \frac{2}{Q} \gamma \partial \phi +  k
\partial \gamma
\end{eqnarray}
where
\begin{eqnarray}
Q & = & \sqrt{\frac{2}{  k}} \nonumber \\
\Delta(\beta,\gamma)& = &(1,0) \nonumber \\
\beta(z) \gamma(w) & \sim & \frac{1}{z-w} \nonumber \\
\phi(w)\phi(z)&  \sim & -\log(z-w)
\end{eqnarray} and the energy momentum tensor of the theory is:
\begin{equation}
T = \beta \partial \gamma -\frac12 (\partial \phi)^2 -\frac{Q}{2}
\partial^2 \phi \,,
\end{equation}
with the central charge given by (\ref{cencha}). The \slr\ vertex
operators are represented by \be \Phi^j_{m,\bar{m}} =
\gamma^{j+m-1} \bar{\gamma}^{j+\bar{m}-1} e^{(j-1)Q \phi} \,.
\label{vertex}
\ee
Correlators were computed in this formalism
first in \cite{Becker:1993at}, by using the screening
charge\footnote{Whenever we say screening charge we imply the
integrated form $\int d^2 z \ {\cal L}$. }
\begin{eqnarray}
{\cal L}_{1} &=& \mu_{1} \beta \bar{\beta} e^{- Q \phi} \,.
\label{beckercharge}
\end{eqnarray}
The two-point function is
\begin{eqnarray}
\langle \Phi^j_{m,\bar{m}} \Phi^{j'}_{m,\bar{m}} \rangle &=&
|z_{12}|^{-4 \Delta_j} \delta_{n+n'} \delta_{w+w'} \left(
\delta(j+j'-1) + R^{NS}(j,m,\bar{m}) \delta(j-j')
 \right)
\label{corr}
\end{eqnarray}
where  $R^{NS}(j,m,\bar{m})$ is  given in (\ref{tpf}).
Similar free field computations can be
performed using a dual screening charge
\cite{Dotsenko:ui,Giribet:2001ft,Hofman:2004ny}
\begin{eqnarray}
{\cal L}_{2} &=& \mu_{2} (\beta \bar{\beta})^{k} e^{-
\frac{2}{Q} \phi} \,,
\label{gncharge}
\end{eqnarray}
and the correlators agree under the identification
\cite{Giribet:2001ft} \be \pi \mu_{2}
\frac{\Gamma(k)}{\Gamma(1-k)} = \left( \pi \mu_{1}
\frac{\Gamma(k^{-1})}{\Gamma(1-k^{-1})} \right)^k \,. \label{mapcoup}
\ee
The vertex operators (\ref{vertex}) correspond to the \slr\
theory. To obtain the primaries of the coset one multiplies them
by the exponential of a free boson, which represents the gauged
coordinate. But  the nontrivial part of any correlator computed is
the same in \slr\ or \slc\, since both interaction terms \bb\ and
\gn\ commute with the gauged current $j^3$.

The free field formalism actually allows  to compute
(\ref{corr}) by inserting both \bb\ and \gn, and only for those
values of $j$ where the anomalous momentum conservation for $\phi$
is satisfied as \be 2(j-1)Q - n_{1}Q -n_{2}\frac{2}{Q} = -Q \,,
\label{amc} \ee
where $n_{1},n_{2}=0,1, \dots $ are the number of
insertions of \bb\ and \gn, respectively\footnote{Note that the
values of $j$ selected by (\ref{amc}) are nothing but $2j=1+n_1+  n_2k$, which correspond to
degenerate representations of the affine \slr\ algebra. \label{degenerate}}. The results are then
analytically continued to arbitrary $j$.

In the limit $\phi \rightarrow \infty$ the potential
drops exponentially and the theory becomes weakly coupled.
In the cigar geometry, the dilaton becomes linear in
$\phi$, which has the interpretation of the radial coordinate away from the tip. Following
\cite{Kazakov:2000pm}, it is easy to see that the
relation between \bb\ and \gn\ is that of a strong-weak coupling
duality. For $k \rightarrow \infty$, \gn\ is supported in the
strong coupling region ($\phi \rightarrow -\infty$) and \bb\ has
support in the weak coupling region. This is consistent with the
fact the \bb\ screening can be obtained from the geometry of the
parent theory, by parameterizing  $AdS_3$ with Poincare
coordinates, and the geometry of the cigar becomes weakly curved
at $k \rightarrow \infty$ since the radius tends to infinity. The
relation is inverted at the opposite limit $k \rightarrow 0$.
Notice that all this is consistent with the fact that $\mu_{1}$
and $\mu_{2}$ are interchanged under $k \leftrightarrow k^{-1}$
as follows from (\ref{mapcoup}).

It was first observed in \cite{FZZ} that correlators computed in
the bosonic \slc\ model coincide with those of the sine-Liouville
model, whose fields are a compact boson $X$ at the radius
$R=\sqrt{2(k+2)}$ of the cigar, and a non-compact one $\phi$ with
background charge. The stress tensor of the theory is
\be
T=
-\frac{1}{2}(\partial X)^2 -\frac{1}{2}(\partial \phi)^2
-\frac{Q}{2} \partial^2 \phi \,, \label{stsl}
\ee
and the central charge is\footnote{This central charge differs by $1$ from the central
charge in  (\ref{cencha}). This corresponds to the addition of a trivial boson as mentioned
before.}\, $c= 2 + \frac{6}{k}$. The \slc\ primaries are
mapped to the following primaries of sine-Liouville:
\be \Phi^j_{m,\bar{m}} =
e^{im\sqrt{\frac{2}{k+2}}X_L + i\bar{m}\sqrt{\frac{2}{k+2}}X_R }
e^{(j-1)Q \phi} \,.
\label{slprimaries}
\ee
The correlation functions are computed in the
Coulomb formalism by using the screening charges
\begin{eqnarray}
{\cal L}_{sl}^{\pm} = \mu_{sl} e^{\pm i \sqrt{\frac{k+2}{2}}
(X_L-X_R)} e^{-\frac{1}{Q}\rho} \,, \label{slscreen}
\end{eqnarray}
which are primaries of $\pm 1$ winding in the compact boson. The
relevant computations can be found in \cite{Baseilhac:1998eq} for
the two-point functions and in \cite{Fukuda:2001jd} for some three point
functions.\footnote{The analytical structure of correlators
computed in \cite{Fukuda:2001jd}, was recently shown in
\cite{Giribet:2004zd} to agree with that obtained in the \slr\
approach for winding-violating processes. } Using \sl\ as
screening charges, the anomalous momentum conservation for a two-point function
is \be 2(j-1)Q  -(n^- + n^+)\frac{1}{Q} = -Q \,,
\label{amcsl} \ee where $n^{\pm}$ is the number of insertions of
\sl.

An important result of \cite{FZZ} is that in the computation of
$N$-point functions, with $N \geqslant 3$, the correlators can violate
winding number by up to $N-2$ units. This result is easily
obtained in the sine-Liouville side, as shown in
\cite{Fukuda:2001jd}, since the integrals to which the correlators
reduce in the Coulomb formalism,  vanish when the
difference $n^- - n^+$ does not have the correct value.
The same result, of course, appears in the \slc\ side, though
through some hard work \cite{Giribet:2001ft, Maldacena:2001km}.
Note that as a consequence, the perturbative computation of a two-point function
requires $n^- = n^+$, and thus an even amount of
insertions of \sl.

In terms of strong-weak coupling duality, the
sine-Liouville \sl\ interaction belongs to the same side of the duality
as the interaction Lagrangian \gn.
The relation between the coupling constants $\mu_{sl}$ and
$\mu_{1}$ was shown in \cite{Giveon:2001up} to be \be \left(
\frac{\pi \mu_{sl}}{k} \right)^{\frac{2}{k}}  = \pi \mu_{1}
\frac{\Gamma(k^{-1})}{\Gamma(1-k^{-1})} \label{mapcoup2} \ee from
which, using (\ref{mapcoup}) it follows that \be \left( \frac{\pi
\mu_{sl}}{k} \right)^{2}= \pi \mu_{2}
\frac{\Gamma(k)}{\Gamma(1-k)}. \ee This quadratic relation between
$\mu_{2}$ and $\mu_{sl}$ is consistent with KPZ scaling, since it
is clear from (\ref{amc}) that any value of $j$ which is  screened
with $n_{2}$ insertions of \gn, can be equally screened with
twice as much insertions of \sl.

\subsection{The supersymmetric case}
The supersymmetric version of the equivalence between the bosonic coset \slc\ and the sine-Liouville
theory, is the celebrated mirror duality between the supersymmetric coset \slc\ and
the $N=2$ Liouville theory. This duality was conjectured in~\cite{Giveon:1999px}.
It  was then shown in \cite{Hori:2001ax} that the
equivalence between the actions of this two $N=2$ theories follows from mirror symmetry,
using the techniques of \cite{Hori:2000kt}. The same equivalence was shown
in  \cite{Tong:2003ik} to follow from  an analysis of the dynamics of domain walls.

In both analyses of \cite{Hori:2001ax} and \cite{Tong:2003ik},  crucial use is made of
the explicit  $N=2$ supersymmetry in the action.
On the other hand, it is easy to see that the computational content of the duality, i.e.,
the identity of the correlators, is exactly the same as that of the bosonic version.
In the supersymmetric coset \slc\ side, the primaries are obtained from the parent susy \slr$_{k}$ model.
In the latter, one can decouple the fermions and shift the level $k \rightarrow k+2$ (see Appendix B),
so that the NS primary states are  the product of a bosonic \slr$_{k+2}$ primary
and the fermionic vacuum. Descending
to the coset involves removing a free $U(1)$ boson, so the correlators reduce
to those of a purely bosonic model at level $k+2$, with the screening charges given by
(\ref{beckercharge}) and (\ref{gncharge}). In the Ramond sector, a coset
primary is obtained by extracting the contribution of the gauged
$U(1)$ from the product of an \slr$_{k+2}$ primary and a spin field
of the decoupled fermions (see e.g.~\cite{Giveon:2003wn}). The computation
reduces also to that of the bosonic case, but the dependence on the quantum number $m$
is now shifted to $m \pm \frac12$, as can be seen in (\ref{tpframond}).

The $N=2$ Liouville is an interacting theory for
a complex chiral super-field $\Phi$, with a Liouville potential.
For some previous works on $N=2$ Liouville
see~\cite{Girardello:1990sh,Kutasov:1990ua,Ahn:2002sx,Nakayama:2004vk}.
The super-field $\Phi$  has  a non-compact real component $\phi$ with background charge,
a compact imaginary component $Y$, and the corresponding fermions. The stress tensor is
\be
T=
-\frac{1}{2}(\partial Y)^2 -\frac{1}{2}(\partial \phi)^2
-\frac{Q}{2} \partial^2 \phi -\frac{1}{2} \psi_{\textsc y} \partial \psi_{\textsc y}
-\frac12 \psi_{\phi} \partial \psi_{\phi}
\label{stntl}
\ee
with a central charge given by  (\ref{cencha}).
The compactness of $Y$ actually allows to define the the following
chiral/anti-chiral superfields in N=(2,2) superspace of coordinates
$(z,\theta^{\pm};\bar{z},\bar{\theta}^{\pm})$:
\be
\Phi^{\pm} = \phi \pm i (Y_L - Y_R) +
i \theta^{\pm} (\tilde{\psi}_{\phi} \mp i \tilde{\psi}_{\textsc y}
) -i \bar{\theta}^{\pm} (\psi_{\phi} \pm i \psi_{\textsc y})
+ i \theta^{\pm} \bar{\theta}^{\pm} F^{\pm} \,.
\ee
Correspondingly we have the following two $N=2$ Liouville chiral interactions:
\be
{\cal L}_{N=2}^{\pm} &=& \mu_{2} \int  d^2 \theta^{\pm} e^{-\frac{1}{Q} \Phi^{\pm}} \nn \\
&=& \frac{\mu_{2}}{Q^2} e^{-\frac{1}{Q} [\phi \pm i (Y_L -  Y_R)  ]}
(\psi_{\phi} \pm i \psi_{\textsc y})(\tilde{\psi}_{\phi} \mp i
\tilde{\psi}_{\textsc y}) \,.
\ee
In the expansion we have set to zero the auxiliary field $F^{\pm}$. Its
presence only contributes contact terms in the correlators, so it can be
ignored \cite{Green:1987qu}.
At this point it is convenient to
bosonize the fermions  as
\be
\frac{\psi_{\phi} \pm i\psi_{\textsc y}}{\sqrt{2}} &=& e^{\pm i H_L} \nn \\
\frac{\tilde{\psi}_{\phi} \pm i \tilde{\psi}_{\textsc y}}{\sqrt{2}} &=& e^{\pm i H_R}
\ee
so that the interaction terms become\footnote{The interaction (\ref{ntlbos}) is
the $N=2$ point in a continuous family of theories studied in \cite{Baseilhac:1998eq}.}
\be
{\cal L}_{N=2}^{\pm} = k \mu_{2}\, e^{-\frac{1}{Q}\phi} e^{\pm i [ \frac{1}{Q}Y_L  + H_L ]}
e^{\mp i [ \frac{1}{Q}Y_R + H_R ]} \,.
\label{ntlbos}
\ee
We will now rotate the two bosons $Y_L$ and $H_L$ as
\be
\sqrt{\frac{k+2}{2}}X_L &=& \frac{1}{Q}Y_L + H_L \nn \\
\sqrt{\frac{k+2}{2}}Z_L &=&  - Y_L + \frac{1}{Q}H_L
\label{rotbos}
\ee
and similarly for $X_R, Z_R$. The two bosons $X_L, Z_L$  commute and are canonically normalized.
Making this change of variables in the
interaction (\ref{ntlbos}), we see that $Z_L$ completely decouples from the
interaction term,
and  \ntl\ becomes identical to the bosonic~\sl, as announced.
This change of variables
is actually the $N=2$ Liouville equivalent of
the chiral rotation in susy \slr\ that allows to decouple the fermions by shifting the level $k$ to $k+2$.
We again arrived to a form of the screening charge without fermions, and the
coefficient of the compact boson in the interaction has been shifted
from~$\frac{1}{Q}= \sqrt{\frac{k}{2}}$~to~$\sqrt{\frac{k+2}{2}}$.
The NS primaries of the model are given by (\ref{slprimaries}). The Ramond sector is treated as
in the \slc\ case with the same result.

By bosonizing the fermions, we have lost
explicit $N=2$ supersymmetry at the level of the action, which was
so important in the approaches of \cite{Hori:2001ax,Tong:2003ik}.
This is not uncommon in conformal field theories, where fermions are typically bosonized in order to compute
correlators\footnote{Explicit  $N=2$ supersymmetry
at the level of the action is not necessary for a conformal field theory to have $N=2$ supersymmetry
in its spectrum and in its chiral algebra.
For example, an $N=2$ minimal model with central charge $c=1$ can be realized through a
free compact boson.
 We thank A.Giveon for comments on this point.}.
In our case it is a signal that the reason for the identity of the correlators may not lie in the
form of the action of the theory (see later).

A dual, non-chiral interaction term is also allowed by the N=2 symmetry
of the N=2 Liouville theory~:
\begin{equation}
\tilde{\cal L}_{N=2} = \tilde{\mu}_2 \int d^4 \theta \ e^{\frac{Q}{2} (\Phi^+ +
  \Phi^-)}.
\end{equation}
We now observe \cite{Nakayama:2004vk} that this screening charge
coincides with the screening charge of the $SL(2,\mathbb{R})/U(1)$
super-coset, i.e. the supersymmetric equivalent of~(\ref{beckercharge}).
Using the same steps of bosonization and field redefinitions one can
reduce this equivalence to the bosonic one already discussed.
Thus this circle of ideas clarifies the fact that the $N=2$ Liouville duality
proposed in~\cite{Ahn:2002sx} is nothing but the bosonic duality of~\cite{FZZ} discussed in the
previous section.

\subsection{Bulk versus localized poles and self-duality}
\label{bulkvslsz}
Poles in the correlators of our model can be either of "bulk" or "localized" type.
Bulk poles correspond
to interactions taking place along the infinite direction of the
(asymptotic) linear dilaton. On the other hand, localized poles are
associated to discrete normalizable states living near the tip of
the cigar\footnote{See \cite{Aharony:2004xn} for a recent
discussion on this double nature of poles in the context of
holographic descriptions of Little String Theories.}.

In particular, in the two-point function (\ref{tpf}), the first two gamma
functions of the numerator have bulk poles, and the last two gamma
functions, with $m,\bar{m}$ dependence, have localized poles. We analyze here the NS two-point function,
the R case being similar.

Let us consider first the bulk poles. The first gamma function has
single poles at values $2j=1,2,\cdots$. From (\ref{amc}), we see that
these values of $j$ are screened by $n_{1}=2j-1$  charges of type
\bb, and no \gn\ charges. In the Coulomb formalism, one first
separates the non-compact  field into $\phi= \phi_0 +
\tilde{\phi}$, and then the integral of the zero mode $\phi_0$
over its infinite volume gives the pole \cite{Goulian:1990qr}.
The second gamma function has poles at $\frac{(2j-1)}{k}=1,2...$.
These values of $j$ are screened in \slc\ by
$n_{2}=\frac{(2j-1)}{k}$  charges of type \gn, and no \bb\
charges. In the sine-Liouville theory, these corresponds to having
$n^+_{sl}=\frac{(2j-1)}{k}$ charges of ${\cal L}_{sl}^{+}$, {\it
and} the same amount of ${\cal L}_{sl}^{-}$. This follows from
(\ref{amcsl}) and the condition $n_{sl}^-=n_{sl}^+$ for two-point functions.

The same phenomenon of having two families of bulk poles occurs in
bosonic Liouville theory, where the two-point function has two
sets of poles, their semi-classical origin being the insertion of
either the Liouville interaction $\mu_{L} e^{-2b \phi}$ or its
dual $\tilde{\mu}_{L} e^{-\frac{2}{b} \phi}$, with a relation
between $\mu_{L}$ and $\tilde{\mu}_{L}$ similar to (\ref{mapcoup})
\cite{Teschner:2001rv}.

The bulk poles in (\ref{tpf}) are simple poles, except at the
level $k=1$. At this level, corresponding to $c=9$, both the first
{\it and} second gamma functions in (\ref{tpf}) have each a pole
at $2j=2,3...$. This is signaled by the fact that the two dual
charges \bb\ and \gn\ become equal for $k=1$, including
$\mu_{1}=\mu_{2}$, so this the self-dual point of the theory.
There is only one single bulk pole left at $k=1$, that comes from
the first gamma function of (\ref{tpf}) at $2j=1$. But this is the
expected result since for this $j$ no screening charges are
needed to satisfy the anomalous momentum conservation (\ref{amc}),
and the single pole comes from the infinite volume of the zero
mode of $\phi$.\footnote{The importance of the self-dual point at
$k=1$ has recently been stressed in \cite{Giribet:2004zd}. The
same self-duality phenomenon, with single poles becoming double
poles, occurs in Liouville at $c=25$ ($b=1$). }

Notice that although one can interpret \bb\ as being a strong-weak
dual to both \gn\ and \sl, the sine-Liouville interaction remains
outside the liaison between the Wakimoto pair \bb\ and \gn, and
this becomes more manifest at the self-dual point $k=1$.

As mentioned, localized poles occur in the third and fourth gamma
functions of the numerator of (\ref{tpf}). They signal the
presence of normalizable bound states in the strong coupling
region associated with the  residue of the singularity. We will
consider here the case $m=\bar{m}=kw/2$, which is the relevant case for
the D-brane analysis. At values of $r =  kw/2-j \geqslant 0$,
corresponding to primaries of \slc\ coming from primaries of \slr,
the fourth gamma function in the numerator of (\ref{tpf}) has a
localized pole. On the other hand, at values of $r =  kw/2-j
<0$ corresponding to primaries of \slc\ which are descendants of \slr,
the two-point function has a zero from a pole in the third gamma function in the
denominator of (\ref{tpf}). We will return to this issue in
sect.~\ref{locbra}.

A qualitative feature that we wish to stress is that,
 for the theories with boundary that we will construct,
{\em the two kinds of poles naturally decouple.}
One-point functions of localized D0-branes will have only poles associated
to the discrete normalizable states, and those of extended
D1-branes will have only "bulk" poles associated to the zero-mode
of the radial coordinate. Intuitively, on the one hand, the
D0-branes are localized near the tip of the cigar, as are the
normalizable bound states, and on the other hand, the D1-branes
stretch along the radial direction and only couple to momentum
modes, thus forbidding the coupling to discrete bound states that
all carry a non-trivial winding charge. In the last case
of D2-branes, we have both type of poles since these non-compact branes
have a induced D0-brane charge localized at the tip of the cigar~\cite{Ribault:2003ss}.

A similar decoupling phenomenon can be observed for the one-point
functions of bosonic (and $N=1$  \cite{Fukuda:2002bv} ) Liouville theory. In this case,
the ZZ one-point functions \cite{Zamolodchikov:2001ah} have no poles and the FZZT
one-point functions \cite{Fateev:2000ik} have the bulk Liouville poles mentioned
above.

\subsection{The conformal bootstrap approach}
We have reviewed above how perturbative calculations with
different screening charges lead to the same correlators in both
theories. The perturbative approach leads to  physical
insights related to the nature of the poles, strong-weak coupling
regimes, etc. Also, in the last years a new powerful approach to
compute correlators in non-rational conformal field theories has
appeared \cite{Teschner:1995yf}. In non-rational
conformal field theories the normalizable states have a continuous
spectrum (in $H_3^+$ they correspond to the continuous
representations of \slr) and appear in the intermediate channels
of the correlators. The new approach consists in assuming that
a general property of the conformal bootstrap, namely the factorization constraints,
can be analytically continued to primary states corresponding to
non-normalizable degenerate operators with discrete spectrum.
This assumption, together with assuming that a  strong-weak
coupling duality is present in the theory (of the type between \bb\ and \gn\ above),
leads to constraints
for two and three point functions which have a unique solution. We
will not review this method here, and we refer the reader to
\cite{Teschner:1995yf,Teschner:2001rv,Fateev:2000ik} for details.

A natural question is what new light can be shed on the
equivalence between $N=2$ Liouville and the susy \slc\ through
these methods. This question is related to the more general issue
of what role  the action or the perturbative screening charges
play in this approach. The method essentially  reduces to a minimum
the dynamical information
needed to solve the theory. It asks as an input two pieces of information:
{\it i)~}the quantum numbers of a degenerate operator, which are given  by
the chiral symmetries of the theory, and {\it ii)}~the value
of certain fixed correlators involving one or two screenings.
As an output we get the value of arbitrary correlators. In the
sine-Liouville/ \slc\ context, this method has been exploited in
\cite{Giveon:2001up} to obtain the relation (\ref{mapcoup2})
between coupling constants $\mu_{1}$ and $\mu_{sl}$.

Now, as shown in \cite{Teschner:1995yf}, it turns out that the second piece
of input, namely, the perturbative fixed correlators,
can be obtained by asking for consistency of the factorization constraints themselves\footnote{We thank V. Schomerus for this crucial comment.}.
This means that the chiral algebra of the model would fully fix
the correlators,
through the family of its degenerate primaries.
The argument holds both for bulk and boundary theories.
In this way, for example, in
\cite{Ponsot:2001gt}, factorization constraints for the boundary $H^+_3$
theory were obtained without introducing a boundary action. In
other words, under certain analyticity assumptions one could in
principle achieve for non-rational conformal field theories, what
is known to hold for rational ones, namely, that the chiral symmetries of
the theory completely fix the correlation functions (under a certain
prescription as to how left and right fields are glued). Notice that
in our theory many factors of the two-point function (\ref{tpf})
can be seen as the result of Fourier-transforming the same object
from a basis of primaries where the \slr\ chiral symmetry is
realized through differential operators (see Appendix
\ref{FT}). So the idea would be to push these symmetry
constraints further to their very end.

Carrying this program in our case, the
\slc\ and $N=2$ Liouville theories  would appear  just as different
realizations of  the same chiral  structure.
The latter would be nothing but the common core of their
respective chiral algebras, affine \slr\ and $N=2, c>3$ Virasoro,
which are  \cite{Dixon:1989cg}
related by a free boson which does not affect the correlators.
In appendix \ref{embedding} we show how an $N=2, c>3$ algebra yields always
an \slr\ algebra by adding a free boson.
Concerning the possible additional "geometrical" information, namely, the
way left and right chiral fields are glued, there is only one known modular invariant with
$N=2, c>3$ spectrum \cite{Eguchi:2004yi,Israel:2004ir}, up to discrete
orbifolds acting on the N=2 charges.

That is the idea behind a central
assumption of our paper, namely, that boundary CFT quantities computed
in $H^+_3$, which descend naturally to \slc, describe also the
boundary theory of $N=2$ Liouville.


\section{The boundary}
\label{boundary}
\subsection*{The bosonic coset}
Recall that for the bosonic coset
we have that the one-point
functions \cite{Ribault:2003ss}
are given in terms of the product of one-point functions for $H_{3}^+$ and
the one-point functions for an auxiliary boson $X$ (at radius
$R=\sqrt{\alpha'k}$)
which we can give the
geometric interpretation of being the angular variable in the coset. The
left and right chiral $U(1)$ quantum numbers of $H_{3}^+$ (labeled $n_H$ and $p_H$)
and $X$ are related as:
\begin{eqnarray}
\left( \frac{n_H+ip_H}{2} ,-\frac{n_H-ip_H}{2}  \right) &=&  \left(\frac{n+kw}{2},
\frac{n-kw}{2} \right)
\end{eqnarray}
where the relative minus sign arises because we gauge
axially.
We then use the factorization of the one-point function:
\begin{eqnarray}
\langle \Phi^{j,coset}_{n,w} \rangle &=& \langle
\Phi^{j,H_3}_{n,w} \rangle  \langle V_{n,w}^X \rangle
\end{eqnarray}
to obtain the one-point function in the coset \cite{Ribault:2003ss}.
Since $X$ has the geometric
interpretation of being the angular variable, Dirichlet conditions on $X$
(i.e. Neumann conditions on the $H_{3}^+$ gauged current) have the interpretation of
branes localized in the angular direction.\footnote{See next paragraph for a
  more rigorous argument based on BRST symmetry.}
Note that we have assumed the ghost contribution to the one-point function
to be trivial. (We can detect non-trivial renormalizations through the Cardy
check.)
Note also that our derivation is only valid for coset primaries that are
associated to {\em primaries} in the parent theory. Coset
primaries associated to descendents in the original model require special
care.
\subsection*{The super-coset}
For the super-coset, we can tell an analogous story. We have that the
one-point functions are given by a product of one-point functions for $H_{3}^+$ at level
$k+2$, for a $U(1)$ associated to the fermions, and for an auxiliary boson $X$ that has
again the interpretation of the angular direction. The one-point functions
factorize, and we have:
\begin{eqnarray}
\langle \Phi^{j,supercoset}_{n,w} \rangle &=& \langle
\Phi^{j,H_3}_{n,w} \rangle
\langle V_{n,w}^X \rangle \langle V_F \rangle.
\end{eqnarray}
We can be more precise about the relationship between the boundary condition
for the various currents. The left and right $N=2$ R-currents are
given in terms of the total currents $J^3$, $\bar{J}^3$ as follows
(see appendix~\ref{characters} for conventions)~:
\begin{equation}
J^{R} = \psi^+ \psi^- + \frac{2J^3}{k} \ \text{and} \quad
\bar{J}^{R} = \tilde{\psi}^+ \tilde{\psi}^-  - \frac{2\bar{J}^3}{k}.
\label{Rcurrentsleftright}
\end{equation}

The {\em A-type} boundary conditions of the N=2 algebra~\cite{Ooguri:1996ck}
are defined through the \emph{twisted} gluing conditions\footnote{All the
gluing conditions in this section correspond to the closed string channel.}:
$J^R =  \bar{J}^R$, $G^{\pm} = i\eta \tilde{G}^{\mp}$, $\eta=\pm 1$ being
the choice of spin structure. Thus we see from~(\ref{Rcurrentsleftright})
--~and the expressions for the supercurrents~-- that the total currents
of the $SL(2,\mathbb{R})$ algebra has to satisfy \emph{untwisted} boundary
conditions: $J^{3,\pm} = - \bar{J}^{3,\pm}$. As we know
from~\cite{Bachas:2000fr,Lee:2001gh,Ponsot:2001gt} this corresponds to $AdS_2$ branes in $AdS_3$.
The axial super-coset has a BRST charge corresponding to the gauge-fixed local
symmetry, whose expression is (see e.g.~\cite{Rhedin:1995um})~:
\begin{equation}
{\cal Q}_{BRST} = \oint \frac{d z}{2i\pi} \left\{ c \, ( J^3 + i\partial X) +
\gamma \, (\psi^3 + \psi^\textsc{x} ) \right\}  -
\oint \frac{d \bar{z}}{2i\pi} \left\{ \tilde{c} \, ( \bar{J}^3 - i\bar{\partial} X) +
\bar{\gamma} \, (\tilde{\psi}^3 - \tilde{\psi}^\textsc{x} ) \right\}.
\end{equation}
Thus the preservation of the BRST current will impose that the extra boson $X$
has the boundary conditions $\partial X = \bar{\partial} X$ in the closed string channel,
i.e. Dirichlet
conditions. The net effect of the $(\beta,\gamma)$ super-ghosts will be to
remove the contributions of the fermions $\psi^3$ associated to $J^3$
and $\psi^\textsc{x}$ associated to $X$, leaving the fermions $\psi^\pm$
with (relative) A-type
boundary conditions. In the cigar these are the D1-branes,
extending to infinity \cite{Fotopoulos:2003vc,Ribault:2003ss}.

The {\em B-type} boundary conditions of the N=2 algebra~\cite{Ooguri:1996ck}
are defined through the \emph{untwisted} gluing conditions:
$J^R =  -\bar{J}^R$, $G^{\pm} = i\eta \tilde{G}^{\pm}$. Using the same
lines of reasoning we find that these boundary conditions corresponds
to \emph{twisted} boundary conditions for the $SL(2,\mathbb{R})$
currents, $J^3=\bar{J}^3,J^{\pm}=-\bar{J}^{\mp}$. These are either $H_2$ branes or $S^2$ of imaginary
radius. In the former case we obtain D2-branes in the cigar, and
in the latter D0-branes localized at the tip of the cigar~\cite{Fotopoulos:2003vc,Ribault:2003ss}.
In both cases the extra field $X$ has to satisfy Neumann boundary conditions.

To summarize, the one-point functions for the super-coset
are as in \cite{Ribault:2003ss} (but, importantly, the basis of Ishibashi
states to which they correspond is a basis of Ishibashi states that
preserves $N=2$ superconformal symmetry and the level is shifted by
two units), with an additional factor corresponding to
two real fermions.

We move on to apply the dictionary above to the particular cases
of D0-, D1- and D2-branes in the supersymmetric $N=2$ theories. We will be
using the quantum number notations traditional for the \slc\ super-coset, for
convenience of comparison with (technically similar) results in the bosonic
coset \cite{Ribault:2003ss}. But it should
always be kept in mind that the construction equally well applies to $N=2$
Liouville theory, since the primaries in one theory can be associated to unique
primaries in its dual and since the characters in both theories are
identical.


\section{D0-branes}
\label{D0NS}

The first type of branes we discuss are the B-type branes
localized at the tip of the
cigar $SL(2,\mathbb{R})/U(1)$ conformal field theory.

\subsection{One-point function for the localized branes}
\label{locbra}

These one point functions  for the D0-branes of N=2 Liouville
are similar to those of the ZZ branes \cite{Zamolodchikov:2001ah} for
bosonic Liouville theory. Their expression is (see Appendix \ref{FT})~:
\be
\langle \Phi_{nw}^j \rangle^{D0}_u &=& \delta_{n,0}
\frac{ \Psi_{u}(j,w) \,\,\,\,\,}{|z-\bar{z}|^{\Delta_{j,w}}}
\label{circleamplitude}
\ee
where
\begin{eqnarray}
\Psi_{u}^{NS}(j,w) &=& k^{-\frac12} (-1)^{uw}
\nu^{\frac{1}{2}-j} \frac{\Gamma(j+\frac{kw}{2}) \Gamma(j-\frac{kw}{2})}
{\Gamma(2j-1) \Gamma(1-\frac{1-2j}{k})} \frac{\sin \frac{\pi}{k} u(2j-1)}
{\sin \frac{\pi}{k} (2j-1)} \nonumber \\
\Psi_{u}^{\nst}(j,w) &=& i^{w} \Psi_{u}^{NS}(j,w)  \nonumber \\
\Psi_{u}^{R^{\pm}}(j,w) &=& k^{-\frac12} (-1)^{uw}
\nu^{\frac{1}{2}-j} \frac{\Gamma(j+\frac{kw}{2}\pm\frac12) \Gamma(j-\frac{kw}{2}\mp \frac12)}
{\Gamma(2j-1) \Gamma(1-\frac{1-2j}{k})} \frac{\sin \frac{\pi}{k} u(2j-1)}
{\sin \frac{\pi}{k} (2j-1)}
\label{opfloc}
\end{eqnarray}
The normalization constant  $k^{-\frac12} (-1)^{uw}$ (and $i^{w}$) is  fixed to satisfy
the Cardy condition (see next section).

The numbers $u=1,2...$ are inherited from the one-point functions
of localized D-branes in $H^+_3$, and they correspond to finite dimensional representations
of \slr\ of spin $j=-\frac{(u-1)}{2}$. The latter  are only a subset of the degenerate
representations of the \slr\ affine algebra\footnote{They correspond  to  $n_2=0 $ in footnote \ref{degenerate}.}.
Remember that in the bosonic and $N=1$ Liouville theory, the localized ZZ branes
\cite{Zamolodchikov:2001ah, Fukuda:2002bv}
have {\it two} quantum numbers, associated to all the degenerate representations of the ($N=1$) Virasoro algebra.
This suggests that  more general solutions of the factorization constraints in \cite{Ponsot:2001gt}
are expected, which in turn would imply a bigger family of D-branes for our $N=2$ model.

For a discrete state with pure winding, the numbers
$j$ and $w$ are correlated as \cite{Israel:2004ir}
\be
j + r &=& \frac{kw}{2}
\label{cosj}
\ee
with $r \in \mathbb{Z}$ for the NS sector and $r \in \mathbb{Z}+\frac12$ for the R sector.
Moreover, the unitarity bound (\ref{unitbound}) for $j$ implies that for every $r$ there is
a unique pair of~$j,w$~such that (\ref{cosj}) holds.

Given a $j$ in the unitary bound (\ref{unitbound}) and satisfying (\ref{cosj}),
the one point functions (\ref{opfloc}) have
poles at values of $r \geq 0, w>0$. The case $r<0$ should be treated differently, since
from the point of view of the parent $H_{3}^+$ --~or $SL(2,\mathbb{R})$~-- theory,
the states with $w>0$ and $w \leqslant 0$ are of very different origin. Indeed
in the former case, the equation~(\ref{cosj}) can be solved with $r \geqslant 0$, hence
those states descend from flowed primaries of a lowest weight representation
$\mathcal{D}_{+}^{j \, , \, w}$ of the $SL(2,\mathbb{R})$ affine
algebra, see~\cite{Henningson:1991jc,Maldacena:2000hw}.
Accordingly the formulae~(\ref{opfloc}) are obtained from descent
of the $H_{3}^+$ ones --~as in~\cite{Ribault:2003ss}~--  valid for primaries
of $H_{3}^+$. The one-point function has simple poles for all these
states, coming form the second Gamma function in the numerator of~(\ref{opfloc}).
On the contrary, in the latter case $w \leqslant 0$, the solution
of~(\ref{cosj}) is solved with $r<0$. Those states, while primaries of the
coset,\footnote{In fact these ``diagonal'' states are obtained from the lowest
weight state of the representation. In a bosonic model they are: $(J_{-1}^{-} )^r |j,j\rangle$.
For the supersymmetric case, see appendix \ref{characters}.} are
\emph{descendents} of the $SL(2,\mathbb{R})$ flowed algebra. To use
nevertheless the formulas of the $H_{3}^+$ branes, we have to use the
isomorphism of representations~: $\mathcal{D}_{j}^{+ \, , \, w} \sim
\mathcal{D}_{j'}^{- \, , \, w-1}$, with $j' = \frac{k+2}{2} -j$. Applying
this mapping on the one-point functions, we find poles coming form
the first Gamma function in the numerators of~(\ref{opfloc}).

\subsection{Cardy computation for the D0-branes}
In this section we will verify that the one-point functions of localized
D-branes obtained in section~\ref{locbra}.
satisfy the Cardy condition relating the open and closed string channels.
We distinguish between three cases for the open string spectrum: NS, R and \nstr.
As usual~\cite{Matsuo:1986vc}, they correspond to NS, \nstr\ and R sectors in
the closed string channel, respectively.
We will show below the computation in detail for the NS/NS case.
For the other cases, we state the result, and defer details to Appendix \ref{ccrn}.

\begin{figure}[!h]
\begin{center}
\epsfig{figure=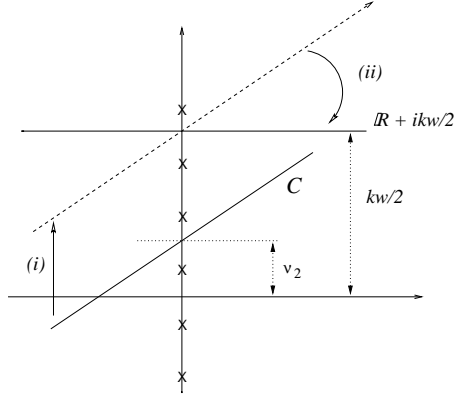, width=60mm}
\caption{Change of contour of integration (case $w>\frac{k \nu_2}{2}$).}
\label{defcont}
\end{center}
\end{figure}

For the annulus partition function of a NS open string stretching between
a brane labeled by $u=1,2...$ and the basic brane with $u'=1$
we consider
\be
Z^{NS}_{u,1}(\tau,\nu) &=& \sum_{r \in \mathbb{Z}} ch_f^{NS} (u,r ; \tau ,\nu ) \nn \\
&=& \frac{\vartheta_3(\tau,\nu)}{\eta(\tau)^3} \sum_{s \in \mathbb{Z}+ \frac12}
\frac{1}{1+yq^s} (q^{\frac{s^2 -su}{k}} y^{\frac{2s-u}{k}} - q^{\frac{s^2
    +su}{k}} y^{\frac{2s+u}{k}}) \,\,.
\label{pfsb}
\ee
We have taken the $N=2$ NS characters (\ref{finchar}) associated to the
$u$-dimensional representations of \slr\ of spin $j= -\frac{(u-1)}{2}$,
and we have summed over the whole spectral flow orbit.
For an open string stretching between two branes with general boundary conditions
$u$ and $u'$, we expect, as in~\cite{Giveon:2001uq,Ponsot:2001gt,Ribault:2003ss},
that the partition function is obtained by summing $Z^{NS}_{u,1}$ over the
irreducible representations appearing in the fusion of the representations
associated to $u$ and $u'$.
Calling the latter ${\bf j}=|j|$ and ${\bf j'}=|j'|$, we have
\be
{\bf j} \otimes {\bf j'} = |{\bf j}-{\bf j'}| \ \oplus \ |{\bf j}-{\bf j'}| +1 \oplus
\cdots \oplus \ {\bf j} + {\bf j'} \,.
\label{fusion}
\ee
In terms of the $u,u'$ indices, this decomposition implies
\be
Z_{u,u'}^{NS}(\tau,\nu) = \sum_{n=0}^{\min(u,u')-1} Z^{NS}_{u+u' -2n -1,1}(\tau,\nu) \,.
\label{uut}
\ee
In order to verify the Cardy condition on these branes, we will perform a
modular transformation of $Z^{NS}_{u,u'}$ to the closed string
channel\footnote{The result for the NS/NS case
follows as a particular case of an identity proved in
\cite{Miki:1989ri,Israel:2004xj}, and the other cases considered below
(\nstr/R, R/\nstr\ and $\mathbb{Z}_p$ orbifold) are variations thereof.
We perform the computation
here for completeness and in order to adapt the
notation to our present purposes.}.
Let us start with the $Z^{NS}_{u,1}$ partition function~(\ref{pfsb}).
We will parameterize $\nu$ as: $\nu = \nu_1 -\tau \nu_2$,
$\nu_{1,2} \in \mathbb{R}$.
The modular transform of
$Z^{NS}_{u,1}(\tau,\nu)$ can be expressed as
\begin{eqnarray}
e^{-i \pi \frac{c}{3} \frac{ \nu^2}{\tau} }  && Z_{u,1}^{NS}(-\frac{1}{\tau},\frac{\nu}{\tau}) =
\frac{\vartheta_3(\tau,\nu)}{\eta(\tau)^3} \nn \\
\times &&
 \frac{1}{2 i \pi}  \left[ \int_{\mathcal{C}_{-\epsilon}}
+ \int_{\mathcal{C}_{+\epsilon}} \right] dZ \
(-i \pi) \, e^{ \pi Z + \frac{2i\pi \tau }{k} Z^2}\,\,  \frac{ \sinh(2 \pi  \frac{Z}{k}u)}
{\cosh (\pi Z)  }  \frac{e^{i\pi  (i\tau Z - \nu  )} }{\cos \pi (i\tau Z-\nu )}
\label{contourint}
\end{eqnarray}
where the parameter $\epsilon$ is infinitesimal and positive.
\begin{figure}[!h]
\begin{center}
\epsfig{figure=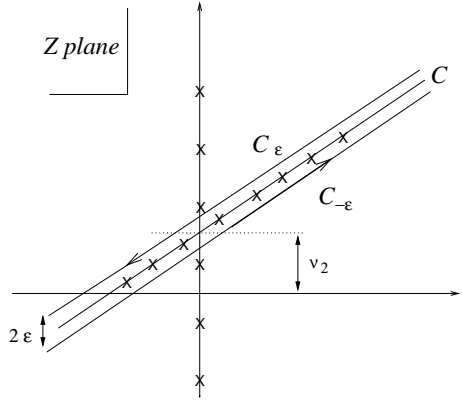, width=60mm}
\caption{Choice of contour of integration (for $\tau_1>0$).}
\label{cont}
\end{center}
\end{figure}
\ni
The contour encircles the line $\mathcal{C}:\ Z =i y / \tau+i\nu_2$,~$y \in
\mathbb{R}$ (see fig.~\ref{cont}).
This identity is valid because  we pick up all the poles of the integrand inside the contour.
These are the zeroes of $\cos \pi (i\tau Z -\nu )$ which
occur at $\nu -i\tau Z  = y + \nu_1 =s \in \mathbb{Z} + 1/2$.
At these poles, the contour integral
yields the residues
\be
 -\frac{2}{\tau}
\frac{e^{  i\pi   (s-\nu)/\tau} \,\, \sin(2 \pi \frac{u}{k}   (s-\nu)/\tau)
  \,\, e^{- i2\pi (s-\nu)^2/(k\tau )}  }
{ e^{i\pi (s -\nu )/\tau} + e^{-i\pi      (s-\nu )/\tau}},
\ee
which lead to the identity (\ref{contourint}).
This identity is valid
as long as there are no poles coming from the $\cosh \pi Z$ factor
on the contour of integration. These special cases occur only
for $\nu_2 \in \mathbb{Z}+1/2$ and we assume that $\nu_2$ does not
take these values.
We now note that we have the
expansions:
\begin{eqnarray}
\frac{e^{i\pi  (i\tau Z - \nu  )} }{2 \cos \pi (i\tau Z-\nu )} &=& \sum_{w=0}^{\infty} (-1)^w
e^{-  2 \pi i w (i\tau Z - \nu )} \ \ \mathrm{for} \ \ | e^{- i 2 \pi (i\tau Z - \nu )} | <1
\label{firstexp}
\end{eqnarray}
and
\begin{eqnarray}
\frac{e^{i\pi  (i\tau Z - \nu  )} }{2 \cos \pi (i\tau Z-\nu )} &=& -\sum_{w=-\infty}^{-1} (-1)^w
e^{-  2 \pi i w (i\tau Z - \nu )} \ \ \mathrm{for} \ \ | e^{+ i 2 \pi (i\tau Z - \nu )} | <1 \,.
\label{secexp}
\end{eqnarray}
The expansion (\ref{firstexp}) is valid in $\mathcal{C}_{+ \epsilon}$
and (\ref{secexp}) is valid in $\mathcal{C}_{- \epsilon}$.
Plugging these expansions into the right hand side of
eq.~(\ref{contourint}) we get
\begin{eqnarray}
e^{-i \pi \frac{c}{3} \frac{ \nu^2}{\tau} }
Z_{u,1}^{NS}(-\frac{1}{\tau},\frac{\nu}{\tau})
= \frac{\vartheta_3(\tau,\nu)}{\eta(\tau)^3}
\sum_{ w \in \mathbb{Z} }
 \int_{\mathcal{C}} dZ \ (-1)^{w} \
\frac{ \sinh(2 \pi \frac{Z}{k} u)} {\cosh(\pi Z)}
 e^{\pi  Z} \ y^w q^{ -i w Z + \frac{Z^2}{k}} \,.
\label{BJ}
\end{eqnarray}
When taking $\epsilon \rightarrow 0$
we added a minus sign to the contour $\mathcal{C}_{+\epsilon}$
and switched its direction.
To obtain the corresponding expression for the general case (\ref{uut}) of $Z^{NS}_{u,u'}$,
we use the identity
\be
\sum_{n=0}^{\min(u,u')-1} \sinh \left(2 \pi \frac{Z}{k} (u+u' -2n -1) \right) =
\frac{\sinh(2 \pi \frac{Z}{k} u) \sinh(2 \pi  \frac{Z}{k} u')}{\sinh(2 \pi  \frac{Z}{k} )}
\label{zzsum}
\ee
so  we get
\begin{eqnarray}
&& e^{-i \pi \frac{c}{3} \frac{ \nu^2}{\tau} }
 Z^{NS}_{u,u'}(-\frac{1}{\tau},\frac{\nu}{\tau})
=  \nn \\
&& \qquad \frac{\vartheta_3(\tau,\nu)}{\eta(\tau)^3} \sum_{ w \in \mathbb{Z} }
 \int_{\mathcal{C}} dZ \ (-1)^{w} \
\frac{ \sinh(2 \pi \frac{Z}{k} u) \sinh(2 \pi \frac{Z}{k} u')} {\cosh(\pi Z) \sinh(2 \pi \frac{Z}{k} )}
 e^{\pi  Z} \ y^w q^{ -i w Z + \frac{Z^2}{k}} \,\,.
\label{BJG}
\end{eqnarray}
Note that the exponent of $q$ is complex.
In order to get a real exponent for $q$ we
will~{\it(i)~}shift the $\mathcal{C}$ contour of integration
by $\frac{iwk}{2}-i\nu_2$, for each term indexed by $w$ and {\it (ii)~}tilt it parallel
to the real axis (see fig.~\ref{defcont}).
The resulting exponent of $q$ will be that needed to
get the characters of the continuous representations.
As we shift the contour we  pick poles which
will make up the contributions of the $N=2$ discrete
representations to the closed string channel amplitude.
No poles are picked when tilting to the real axis. So
the modular transform of $Z^{NS}_{u,u'}$ is decomposed into
\begin{eqnarray}
e^{-i \pi \frac{c}{3} \frac{ \nu^2}{\tau} }
Z_{u,u'}^{NS}(-\frac{1}{\tau},\frac{\nu}{\tau}) =
Z_{u,u'}^{NS,c} +
Z_{u,u'}^{NS,d} \,.
\label{BJDIV}
\end{eqnarray}
The first term comes from the continuous integral, which after shifting and tilting the contour, becomes
\begin{eqnarray}
Z_{u,u'}^{NS,c} =
\sum_{ w \in \mathbb{Z} }
 \int_{-\infty}^{+\infty} \!\!\!\! \!\!\!\! dP
\frac{  e^{\pi  (\frac{iwk}{2} +P)}
 \sinh(2 \pi  \frac{P}{k} u ) \sinh(2 \pi  \frac{P}{k} u' )}
{(-1)^{w(u-u')} \cosh(\pi (P+i\frac{wk}{2})) \sinh(2 \pi  \frac{P}{k}  )} &&\\
&& \!\!\!\! \!\!\!\! \times \ \ q^{\frac{P^2+ (wk/2)^2}{k}}y^w
\frac{\vartheta_3(\tau,\nu)}{\eta(\tau)^3} \,.\nn
\end{eqnarray}
In the last factor we get a continuous character $ch_c^{NS} (P,\frac{wk}{2} ; \tau,\nu)$
(see (\ref{carcon1})),
corresponding to a pure winding state with $m=\frac{kw}{2}$, as expected.
And since $ch_c (P,\frac{wk}{2}; \tau,\nu)$ is even in $P$ we can rewrite this as
\begin{eqnarray}
Z_{u,u'}^{NS,c} = \sum_{ w \in \mathbb{Z} }
 \int_{0}^{+\infty} \!\!\! \! \!\!\!\! dP
\frac{   2 \sinh(2\pi P) \sinh(2 \pi  \frac{P}{k} u ) \sinh(2 \pi  \frac{P}{k} u' ) }
{(-1)^{w(u-u')} [\cosh(2 \pi P) \!+\! \cos(\pi k w)] \sinh(2 \pi  \frac{P}{k} ) }
  ch_c^{NS} (P,\frac{wk}{2} ; \tau,\nu) \,. \nn \\
\label{D0cont}
\end{eqnarray}
This expression is equal to
\begin{eqnarray}
Z_{u,u'}^{NS,c} = \sum_{ w \in \mathbb{Z} }
 \int_{0}^{+\infty} \!\!\! dP \
 \Psi_{u}\left(\frac12 -iP,-w \right)\Psi_{u'}\left(\frac12 +iP,w \right)
\  ch_c^{NS} (P,\frac{wk}{2} ; \tau,\nu)
\end{eqnarray}
so we get the expected continuous spectrum from the overlap between D0 boundary
states\footnote{Note that in the "out" boundary state
we take  the opposite $U(1)$ charge, as follows from
CPT conjugation \cite{Recknagel:1997sb}. }.

The poles picked up
arise from the zeroes of the factor $\cosh \pi Z$ in (\ref{BJG}), at values of $Z$ such that
$Z=is$ with $s \in \mathbb{Z} + \frac12$.
Note that for irrational $k$ the shifted contour will never
fall on a pole, and we will assume this to be the case.
The poles will contribute with positive sign for $s > \nu_2$ and with negative sign for
$s < \nu_2$. Let us consider the former case. A given $s \in \mathbb{Z} +
\frac12$ will give a pole contribution for all the terms
in (\ref{BJ}) with $w \geqslant w_s$, where $w_s$ is the only integer satisfying
\be
w_s > \frac{2s}{k} > w_s -1 \,.
\label{ws}
\ee
For poles corresponding to $s < \nu_2$, we get contributions for all the terms in
(\ref{BJ}) with $w \leqslant w_s -1$, where $w_s$ is also defined by (\ref{ws}).
Summing all these residues we get
\begin{eqnarray}
Z_{u,u'}^{NS,d} = -2 \frac{\vartheta_3(\tau,\nu)}{\eta(\tau)^3} \left[ \sum_{s>\nu_2}
\sum_{w=w_s}^{\infty}-\sum_{s<\nu_2} \sum_{w= -\infty}^{w_s-1} \right]
\frac{\sin \left( 2 \pi \frac{s}{k}u \right) \sin \left( 2 \pi \frac{s}{k}u' \right)}
{(-1)^w \sin \left( 2 \pi \frac{s}{k} \right)}
\ y^w q^{sw-\frac{s^2}{k}} \,.
\end{eqnarray}
Noting that $|yq^{s}|=e^{-2\pi \tau_2(s-\nu_2)}$ is smaller (bigger) than $1$
for $s>\nu_2$ ($s<\nu_2$), we can sum on $w$ for every $s$ to get
\begin{eqnarray}
Z_{u,u'}^{NS,d} = -2  \sum_{s \in \mathbb{Z} + \frac12}
(-1)^{w_s}
\frac{\sin \left( 2 \pi \frac{s}{k}u \right) \sin \left( 2 \pi \frac{s}{k}u' \right)}
{\sin \left( 2 \pi \frac{s}{k} \right)}
\
 \frac{y^{w_s}
q^{sw_s-\frac{s^2}{k}}}{1 + yq^s}
\frac{\vartheta_3(\tau,\nu)}{\eta(\tau)^3} \,.
\label{sumdisc}
\end{eqnarray}
This result can be recast in a more transparent fashion as follows. Calling
$s=r+\frac12$ with $r \in \mathbb{Z}$, a character of the $N=2$ NS discrete
representations corresponding to a pure winding mode $J^3_0=j+r=\frac{kw}{2}$
is given by (see (\ref{cardis1}))
\be
ch_d^{NS} (j,r; \tau ,\nu )=   \frac{ y^{w} q^{sw-\frac{s^2}{k}}  }{1+yq^s}
\frac{\vartheta_3(\tau,\nu )}{\eta(\tau)^3} \,.
\ee
Now, for every $r \in \mathbb{Z}$ there is {\it only one} value of $w$, such
that $j=-r + \frac{kw}{2}$ lies in the improved unitary bound
(\ref{unitbound}).
This is exactly the value of $w$ fixed by the condition (\ref{ws}).
Moreover, {\it all} possible representations corresponding to pure winding
states and such that $j$ lies inside the unitary bound are covered by taking
all $r \in \mathbb{Z}$ and fixing
$w_s$ as in (\ref{ws}). Calling $j_r$ the spin of the representation
associated to each $r$, then (\ref{sumdisc}) is equal to
\begin{eqnarray}
Z_{u,u'}^{NS,d} =  \sum_{r \in \mathbb{Z} }
(-1)^{w_r(u-u')}
\frac{2 \sin \left( \frac{ \pi}{k} (2j_r -1)u\right) \sin \left( \frac{ \pi}{k} (2j_r -1)u'\right)}
{\sin \left( \frac{ \pi}{k} (2j_r -1)\right)}
\ ch_d^{NS} (j_r,r ;  \tau ,\nu )
\label{sumdisc2}
\end{eqnarray}
where $w_s$ has been renamed $w_r$.

As shown in section \ref{locbra}, the product $\Psi_{u}(-j+1,-w) \Psi_{u'}(j,w)$
has a single pole for every discrete, pure-winding state.
It is natural to expect the discrete part of the annulus amplitude (in the
closed string channel) to be given by
the residue of this pole. This is indeed the case, and one can check that
(\ref{sumdisc2}) is equal to
\begin{eqnarray}
Z_{u,u'}^{NS,d} =
2 \pi \sum_{r \in \mathbb{Z} } Res
\left[ \Psi_u^{NS}(-j_r +1,-w_r)\Psi_{u'}^{NS}(j_r ,w_r) \right]
\ ch_d^{NS} (j_r,r ;  \tau ,\nu ) \,,
\end{eqnarray}
where the residue is computed when considering the bracketed expression as an
analytic function of~$j$. More details about this last step are
given in section ~\ref{D2} that treats D2-branes.
We have thus verified the Cardy consistency condition
for the D0 branes in the NS/NS sector.

The computation for the R/\nstr\ and \nstr/R cases is basically the same,
{\it mutatis mutandi}. We state here the results and provide the details
of the computation in Appendix \ref{ccrn} for the interested reader.
For the open string sector we take the partition functions
\be
Z^{R}_{u,1}(\tau,\nu) &=& \sum_{r \in \mathbb{Z}+ \frac12}
ch_f^{R} (u,r ; \tau ,\nu ) \,,
\label{pfsbRr}
\ee
\be
Z^{\nst}_{u,1}(\tau,\nu) &=& \sum_{r \in \mathbb{Z}} ch_f^{\nst} (u,r ; \tau ,\nu )\,,
\label{pfsbNSTr}
\ee
and  $Z^{R/\nst}_{u,u'}(\tau,\nu)$ is given by  a sum as in (\ref{uut}).
The modular transforms are
\begin{eqnarray}
&& e^{-i \pi \frac{c}{3} \frac{ \nu^2}{\tau} }
Z_{u,u'}^{R}(-\frac{1}{\tau},\frac{\nu}{\tau}) =
\label{BJDIVRr} \\
&& \qquad \sum_{ w \in \mathbb{Z} }
 \int_{0}^{+\infty} \!\!\! dP \
 \Psi_{u}^{\nst}\left(\frac12 -iP,-w \right)\Psi_{u'}^{\nst}\left(\frac12 +iP,w \right)
\  ch_c^{\widetilde{NS}} (P,\frac{wk}{2} ; \tau,\nu) \nn \\
 && \qquad \qquad
+ 2 \pi \sum_{r \in \mathbb{Z} } Res \left[
\Psi_u^{\nst}(-j_r +1,-w_r)\Psi_{u'}^{\nst}(j_r ,w_r) \right]
\ ch_d^{\widetilde{NS}} (j_r,r ;  \tau ,\nu ) \,, \nn
\end{eqnarray}

\begin{eqnarray}
&& e^{-i \pi \frac{c}{3} \frac{ \nu^2}{\tau} }
Z_{u,u'}^{\nst}(-\frac{1}{\tau},\frac{\nu}{\tau}) = \label{poprr} \\
&& \qquad \sum_{ w \in \mathbb{Z} }
 \int_{0}^{+\infty} \!\!\! dP \
 \Psi_{u}^{R}\left(\frac12 -iP,-w \right)\Psi_{u'}^{R}\left(\frac12 +iP,w \right)
\  ch_c^{R} (P,\frac{wk}{2} ; \tau,\nu) \nn \\
&&  \qquad \qquad
+2 \pi \sum_{r \in \mathbb{Z} + \frac12} Res \left[
  \Psi_u^{R}(-j_r +1,-w_r)
\Psi_{u'}^{R}(j_r ,w_r) \right]
\ ch_d^{R} (j_r,r ;  \tau ,\nu ) \,,  \nn
\end{eqnarray}
where $j_r$ and $w_r$ are defined in the same way as for the NS/NS case, but notice that
in the the R case $r$ is half-integer. In both (\ref{BJDIVRr}) and (\ref{poprr}) the residues
are again computed considering the bracketed expression as a function of $j$.

\vskip .5cm
Some comments are in order.
\begin{itemize}
\item Although the Cardy condition holds for arbitrary boundary conditions $u,u'$,
there are reasons to argue that only the $u=u'=1$ case corresponds
to physical D-branes. Firstly, for $u,u' \neq 1$ the open string
partition function is built from {\it non-unitary} $N=2$ representations.
An alternative argument was given in \cite{Ribault:2003ss}, based
on comparing higher values of $u$ to expectations for the physics of
coinciding single D0-branes.

\item From the case of the D0 branes we draw an important lesson,
which will remain valid for the D1 and D2 branes. Although the open string partition
function is built out of $N=2$ characters, the product of the one point
functions in the closed string channel is  the same as that of the {\it bosonic} 2D black hole
studied in \cite{Ribault:2003ss} at level $k+2$. Using
identities developed in \cite{Israel:2004xj} (see also \cite{Bakas:1991fs}) to relate sums of characters of $N=2$
representations to characters of a bosonic \slc\ model, one can expand
the partition function (\ref{pfsb}) of a string stretching between a $u$-brane and a basic brane as
\be
\label{pfexpanded}
Z_{u,1}^{NS}(\tau ,\nu ) = \frac{1}{\eta(\tau)} \sum_{n,r \in \mathbb{Z}}
z^{\frac{(j+r)}{k/2} +n} q^{\frac{k/2}{k+2}(\frac{(j+r)}{k/2} + n)^2}
\left[ \lambda_{j,r-n}(\tau) - \lambda_{-j+1,r-n-u}(\tau) \right]
\ee
where $j=-\frac{(u-1)}{2}$ and
\be
\lambda_{j, r} (\tau ) = \eta(\tau )^{-2} q^{-\frac{(j-\frac{1}{2})^2}{k} +\frac{(j+r)^2}{k+2}}
\sum_{s=0}^{\infty} (-1)^s q^{\frac{1}{2} s(s+2r+1)}
\ee
are the characters of the bosonic coset \slc\ descending from bosonic \slr\
primaries with $J^3_0=j+r$.
On the other hand, the open string partition function of a bosonic open string stretching
between similar D-branes  in the bosonic cigar background is given by \cite{Ribault:2003ss}
\be
\label{pfbosonicexpanded}
Z_{u,1}^{bosonic}(\tau ,\nu ) = \sum_{r \in \mathbb{Z}} \left[ \lambda_{j,r}(\tau) - \lambda_{-j+1,r}(\tau) \right] \,.
\ee
Notice that the partition functions (\ref{pfexpanded}) and (\ref{pfbosonicexpanded}) differ by
characters of a $U(1)$ boson. This  is the $U(1)$ R-current
of $N=2$, who is responsible for the extension
of the bosonic \slc\ algebra into $N=2$ \cite{Dixon:1989cg}. It is a free boson, and it
is coupled in a way that is of mild consequence to the modular matrix and to
the one-point functions.

\end{itemize}

\vskip 0.8cm


\section{D0 branes in a $\zi_p$ orbifold}
\label{orbifold}
Before proceeding to the D1 branes, we will discuss a natural and interesting generalization
of the D0 branes considered in the previous section.

In the annulus partition function (\ref{pfsb})
we summed over the whole infinite spectral flow orbit,
and this was essential in order to obtain a discrete spectrum of
$U(1)$ charges in the closed string channel. But a consistent result is also
obtained if the sum is taken with jumps of $p$ units of spectral flow.
As we will see, this correspond to D0 branes living in a  $\zi_p$ orbifold of
the original background, and in this framework we will be able to  connect
with previous works on boundary $N=2$ Liouville theory
\cite{Eguchi:2003ik,McGreevy:2003dn, Ahn:2003tt}.

The orbifold  acts freely as a shift of $2\pi/p$ on the angular direction of the cigar.
As for a compact free boson, we expect that the orbifold theory is equivalent
to the original one with the radius  divided by $p$. This implies a spectrum of
charges given by
\be
J^3_0+\bar{J}^3_0=m+\bar{m}=\frac{kw}{p}\ , \quad
J^3_0-\bar{J}^3_0=m-\bar{m}= pn.
\label{specmp}
\ee
Moreover, since the (infinite) volume of the target space decreases now by a
factor $1/p$, we expect that the
one point functions should be renormalized by $\sqrt{1/p}$. For $k$ integer
one can mod out the theory by $\zi_k$ (see below), and we get the
spectrum of single cover of the vector coset $SL(2,\mathbb{R})/U(1)$~\cite{Israel:2003ry},
which has the target space geometry of the trumpet, at first order in $1/k$.

By applying the same method of descent from $H_{3}^+$ we can compute the
one-point function for this orbifold.
For simplicity, we will study here the NS/NS case, with $u=u'=1$,
and we omit the NS/$u$/$u'$ labels..
The other sectors are  similar.

We obtain the following
one-point function for the D0-branes~:
\begin{eqnarray}
\Psi_{p,a}(j,w) &=& \frac{1}{\sqrt{pk}} e^{-2\pi i a \frac{w}{p}}
\nu^{\frac{1}{2}-j} \frac{\Gamma(j+\frac{kw}{2p}) \Gamma(j-\frac{kw}{2p})}
{\Gamma(2j-1) \Gamma(1-\frac{1-2j}{k})}
\label{opflocorbi}
\end{eqnarray}
where $a \in \mathbb{Z}_p$ is an additional
quantum number of these D-branes and the phase
$e^{-2\pi i a \frac{w}{p}}$
is fixed by the Cardy condition.

Let us start with the open string partition function for an open string stretching
between two such  D0-branes, for which we consider
\be
Z_{p;a,a'}(\tau,\nu) &=& \sum_{r \in  \mathbb{Z}p + a-a' } ch_f (1,r  ; \tau ,\nu ) \nn \\
&=& \frac{\vartheta_3(\tau,\nu)}{\eta(\tau)^3}
\left[ \sum_{s \in \mathbb{Z}p + a -a'+ \frac12}
\frac{q^{\frac{s^2 -s}{k}} y^{\frac{2s-1}{k}}}{1+yq^s}
- \sum_{s \in \mathbb{Z}p -1 + a -a'+ \frac12} \frac{q^{\frac{s^2 +s}{k}}
y^{\frac{2s+1}{k}}}{1+yq^s} \right] \,.
\label{pforb}
\ee
We have summed over all the representations of
the spectral flow with $p$-jumps,
with the starting point
$r=a-a'$ coming from the additional parameters of these D0-branes.

In order to check the Cardy condition, we will perform
a modular transformation of $Z_{p;a,a'}$ to the closed string channel.
The computation is very similar to that of the previous section, so we will
indicate only the major steps. We start with
\begin{eqnarray}
\label{contourintorbifold}
e^{-i \pi \frac{c}{3} \frac{ \nu^2}{\tau} }
&& Z_{p;a,a'}(-\frac{1}{\tau},\frac{\nu}{\tau}) =
\frac{\vartheta_3(\tau,\nu)}{\eta(\tau)^3} \, \times
\left. \phantom{\frac{x^{2^{2^2}}}{x_{2_{2_2}}}} \right. \\
&&
\left\{  \frac{1}{2 i \pi}  \left[ \int_{\mathcal{C}_{-\epsilon}}
+ \int_{\mathcal{C}_{+\epsilon}} \right] dZ \
\frac{(-i \pi)}{p} \,
\frac{e^{ \pi Z +2 \pi \frac{Z}{k} + \frac{2i\pi \tau }{k} Z^2}} {\cosh (\pi Z)  }
\frac{e^{-i\pi  (\frac{\nu - i\tau Z -1/2 -(a-a')}{p} +\frac12) } }{2 \cos \pi (
\frac{\nu - i\tau Z -1/2 -(a-a')}{p} +\frac12)} \right. \nn \\
-&& \frac{1}{2 i \pi}  \left[ \int_{\mathcal{C}_{-\epsilon}} +
\left. \int_{\mathcal{C}_{+\epsilon}} \right] dZ \
\frac{(-i \pi)}{p} \,
\frac{e^{ \pi Z -2 \pi \frac{Z}{k} + \frac{2i\pi \tau }{k} Z^2}} {\cosh (\pi Z)  }
\frac{e^{-i\pi  (\frac{\nu - i\tau Z +1/2 -(a-a')}{p} +\frac12) } }{2 \cos \pi (
\frac{\nu - i\tau Z +1/2 -(a-a')}{p} +\frac12)} \right\} \,. \nonumber
\end{eqnarray}
The contour of integration is the same as in the previous section, but now the
last factor of the integrand has poles at values $\nu - i\tau Z = p\mathbb{Z} + a -a'+ \frac12$
in the first term, and at
$\nu - i\tau Z = p\mathbb{Z} + a -a' - \frac{1}{2}$ for the second.
Expanding the last factor of (\ref{contourintorbifold}) as in (\ref{firstexp})-(\ref{secexp}),
and taking $\epsilon \rightarrow 0$ in the contours $\mathcal{C}_{\pm\epsilon}$,
 yields,
\begin{eqnarray}
&& e^{-i \pi \frac{c}{3} \frac{ \nu^2}{\tau} }
 Z_{p;a,a'}(-\frac{1}{\tau},\frac{\nu}{\tau})
=  \nn \\
&& \qquad \frac{\vartheta_3(\tau,\nu)}{\eta(\tau)^3} \sum_{ w \in \mathbb{Z} }
 \int_{\mathcal{C}} dZ \ e^{-2i\pi  (a -a') \frac{w}{p}}
\frac{ \sinh (  \pi  ( \frac{2Z}{k} -\frac{iw}{p}) ) }{p \cosh(\pi Z) }
 e^{\pi  Z} \ y^{\frac{w}{p}} q^{ -i \frac{wZ}{p} + \frac{Z^2}{k}} \,.
\label{BJORBI}
\end{eqnarray}
In order to express (\ref{BJORBI}) as a sum over characters
we shift the contour in each term by an amount of $\frac{iwk}{2p} - i\nu_2$.
As in the previous section, we pick poles for $Z = is$,
with $s \in \zi + \frac12$, corresponding to the discrete representations.
Defining $w_s \in \zi$ as
\be
\label{contshiftp}
w_s > \frac{2ps}{k} > w_s -1 \,\,,
\ee
the poles with $s > \nu_2$ ($s < \nu_2$) are picked with positive
(negative) sign, by all the values $w \geq w_s$ ($w < w_s$) in (\ref{BJORBI}).
The sum of  all
the poles is
\be
&& -2 \frac{\vartheta_3(\tau,\nu)}{\eta(\tau)^3} \left[ \sum_{s>\nu_2}
\sum_{w=w_s}^{\infty}-\sum_{s<\nu_2} \sum_{-\infty}^{w_s-1} \right]
e^{-2\pi i (a-a') \frac{w}{p}}
 \sin \left( 2 \pi \frac{s}{k} - \frac{\pi w}{p}\right)
\ y^{\frac{w}{p}} q^{\frac{sw}{p}-\frac{s^2}{k}} \\
&=&
-2 \frac{\vartheta_3(\tau,\nu)}{\eta(\tau)^3}  \sum_{s \in \zi + \frac12}
\sum_{n \in \zi_p}
e^{-2\pi i (a-a') \frac{w_s + n}{p}}
 \sin \left( 2 \pi \frac{s}{k} - \pi\frac{(w_s + n)}{p}\right)
\ \frac{y^{\frac{w_s + n}{p}} q^{\frac{s(w_s +n)}{p}-\frac{s^2}{k}}}{1+yq^s} \,. \nn
\ee
Collecting the continuous and discrete contributions, we get finally
\begin{eqnarray}
&& e^{-i \pi \frac{c}{3} \frac{ \nu^2}{\tau} }
Z_{p;a,a'}(-\frac{1}{\tau},\frac{\nu}{\tau}) =
\label{cardyorb} \\
&& \qquad \sum_{ w \in \mathbb{Z} }
 \int_{0}^{+\infty} \!\!\! dP \
 \Psi_{p,a'}\left(\frac{1}{2} -iP,-w \right)\Psi_{p,a}\left(\frac{1}{2} +iP,w \right)
\  ch_c^{NS} (P,\frac{kw}{2p} ; \tau,\nu) \nn \\
&& \qquad
+2 \pi \sum_{r \in \mathbb{Z} } \sum_{n \in \zi_p} Res \left[
 \Psi_{p,a'}(-j_{r,n} +1,-w_s-n)\Psi_{p,a}(j_{r,n} ,w_s+n) \right]
\ ch_d^{NS} (j_{r,n},r ;  \tau ,\nu ) \,, \nn
\end{eqnarray}
where as usual the residue is computed when considering the bracketed
expression as a function of~$j$. The spins $j_{r,n}$ of the
discrete representations are defined through
\be
j_{r,n}+ r = \frac{k(w_s +n)}{2p}
\ee
where $s=r+\frac12$, and $j_{r,n}$ satisfies the unitary bound
(\ref{unitbound}). Moreover, from (\ref{contshiftp})
it follows that the width $k/2$ of the unitary bound (\ref{unitbound})
is sliced into $p$ intervals, and for each $n \in \zi_p$ we have
\be
\frac12 + \frac{kn}{2p} < j_{r,n} < \frac{k(n+1)}{2p} + \frac12 \,.
\ee
Finally,
note that the choice of $a$ in the open string channel,
which is a statement about the {\it spectrum}, becomes a {\it phase} in the
closed string channel.

\subsection*{The $\zi_k$ and $\zi_{\infty}$ cases}
In this context, we can now construe previous studies of D-branes in $N=2$
 Liouville theory as particular cases of the orbifold discussed here.

In \cite{Eguchi:2003ik} D-branes were built in a $N=2$
model with rational central charge $c=3 + \frac{6K}{N}$, with $K,N$ positive integers,
and the D0-brane open string partition function was taken
with jumps of $N$ spectral flow units (for $u=u'=1$).
This corresponds to level $k = \frac{N}{K}$ and $p=N$ in our case.
In particular, for $K=1$, the asymptotic radius shrinks from
$R=\sqrt{k \alpha'}$ to $\frac{R}{k}= \frac{\alpha'}{R}$,
which is the T-dual radius\footnote{In the case
of rational $k$ there will be additional discrete terms in the closed string channel partition
function, coming from poles falling exactly on the displaced contour of section \ref{D0NS}.
But taking them into account does not change the orbifold picture.}. Indeed, notice that for
$p=k$, the spectrum of pure winding modes in (\ref{specmp}) becomes  pure
momentum in the T-dual picture.
Note that we strengthen some of the results in  \cite{Eguchi:2003ik},
since our construction is explicitly based on branes which have been checked
to satisfy factorization constraints in the parent theory \cite{Ponsot:2001gt}. Moreover, our
results our based on a systematic analysis of a solution to the branes
preserving the full chiral algebra in the parent theory, and do not depend on
the central charge being rational. Apart from these bonuses, we also clarified the
connection to the semi-classical geometrical (mirror) cigar-picture.

Another interesting case to consider is the limit $p \rightarrow \infty$. This
corresponds to the cigar radius shrinking to zero, so that the spectrum
(\ref{specmp}) of $m$ for pure winding states  becomes continuous,
thus yielding a continuum of R-charges in the closed string channel.
As shown in~\cite{Israel:2003ry} it corresponds to the vector gauging of the
\emph{universal cover} of $SL(2,\mathbb{R})$.

At the level of the Cardy computation, naively taking the limit $p \rightarrow \infty$
in (\ref{cardyorb}) gives zero in the closed string channel, since
$\Psi_{p,a}(j,w)$ goes to zero, see~(\ref{opflocorbi}).
The correct way of proceeding is to notice that equation (\ref{BJORBI}) becomes a Riemann sum,
which leads to an integral over $t=\frac{w}{p} \in \mathbb{R}$. We can moreover
express $t=x + g, \ g \in \mathbb{Z}, \ x \in [0,1)$. The sum over $w$ in~(\ref{BJORBI})
becomes
\be
\frac{1}{p} \sum_{w}  \longrightarrow \int_{-\infty}^{+\infty} dt \longrightarrow
\int_{0}^{1}dx \sum_{g \in \mathbb{Z}} \,.
\label{chain}
\ee
One can show that the resulting expression for the closed string channel
is equivalent to starting with
\be
Z_{1,1}(x;\tau,\nu)  &=& \sum_{r \in  \mathbb{Z} } e^{2 \pi i x r} ch_f^{NS} (1,r  ; \tau ,\nu ) \,,
\ee
and computing
\be
\int_{0}^{1}dx \ e^{- 2 \pi i x d } e^{-i \pi \frac{c}{3} \frac{ \nu^2}{\tau} }
Z_{1,1}(x, -\frac{1}{\tau},\frac{\nu}{\tau}) \,,
\label{iden}
\ee
where $d \in \mathbb{Z}$  depends on the precise way the limit is taken in (\ref{chain}).
The computation of~(\ref{iden}) itself can be performed with an identity proved in
\cite{Miki:1989ri,Israel:2004xj}, to which we refer for details. The resulting expression
can be found in \cite{Ahn:2003tt}.


\section{D1 branes}
\label{D1}
In this section,
we check the relative Cardy condition for D1-branes in the $N=2$
Liouville/supersymmetric coset conformal field theory.
In this and the next sections, we will work in the $NS$ sector.
The other sectors are
 obtained straightforwardly.
Following the general logic, we assume that the
one-point functions for closed string primaries in the presence of a D1-brane
labeled by the parameters $(r,\theta_0)$
is given by (see Appendix B):
\be
\langle \Phi_{nw}^{j} (z,\bar{z}) \rangle^{D1}_{r,\theta_0}
= \delta_{w,0}\ \frac{\Psi_{r,\theta_0} (j,n) }{|z-\bar{z}|^{\Delta_{j,n}}} \ ,
\ee
with
\be
\Psi_{r,\theta_0} (j,n) =
{\cal N}_1
e^{i n \theta_0}
\left\{ e^{- r (-2j+1)} + (-1)^n e^{ r (-2j+1)} \right\}
\nu^{\frac{1}{2}-j}
\frac{\Gamma(-2j+1)\Gamma(1+\frac{1-2j}{k})}{
\Gamma(-j+1+\frac{n}{2})\Gamma(-j+1-\frac{n}{2})} \nn \\
\ee
We put a normalization factor ${\cal N}_1$ up front that will be fixed
during the Cardy computation.

Note that the only poles in the one-point function originate from infinite
volume divergences (i.e. they are bulk poles).
Thus the only contributions
to the closed string channel amplitude between two D1-branes
originate from continuous representations (unlike the D0- and D2-brane
computations where poles associated to discrete representations require
special attention). The closed string channel amplitude will only contain
contributions proportional to the continuous character:
\begin{eqnarray}
ch_{c}^{NS} (P,\frac{n}{2} ; -\frac{1}{\tau}, \frac{\nu}{ \tau} ) &=&
\tilde{q}^{\frac{P^2}{k} + \frac{n^2}{4k}} \tilde{y}^{\frac{n}{k}}
\frac{\vartheta_3
(-\frac{1}{\tau},\frac{\nu}{\tau} )}{ \eta^3 (-\frac{1}{\tau} )} .
\end{eqnarray}
We compute the closed string channel amplitude between two branes with
identical boundary conditions both labeled by $(r,\theta_0)$, for
simplicity. The computation can then easily be generalized to include
the case of differing boundary conditions, following
\cite{Ribault:2002ti}. For our case, we find the
partition function:
\begin{eqnarray}
Z_{r,\theta_0}^{D1} &=& \int_0^{\infty} dP \sum_{n \in  \zi}
\Psi_{r,\theta_0} (P,n)
\Psi_{r,\theta_0} (-P,-n) \
ch^{NS}_c \left( P,\frac{n}{2} ; -\frac{1}{\tau} , \frac{\nu}{\tau} \right) \\
&=&  \frac{4 {\cal N}_1^2}{k} \int_0^{\infty} dP  \ \frac{1}{\sinh 2 \pi P
\sinh \frac{2 \pi P}{k}} \nonumber \\
& & \left( \sum_{n \in  2Z}   \cos^2 2 r P \cosh^2 \pi P\ ch^{NS}_c(P,\frac{n}{2})
+  \sum_{n \in  2Z+1}   \sin^2 2 r P \sinh^2 \pi P \ ch^{NS}_c(P,\frac{n}{2})
\right).
\nonumber
\end{eqnarray}
We modular transform the characters to obtain the annulus amplitude suitable
for interpretation in the open string channel:
\begin{eqnarray}
e^{- \frac{c i \pi \nu^2}{3 \tau}} Z_{r,\theta_0}^{D1}
&=& \frac{8 {\cal N}_1^2}{k} \int_0^{\infty} dP   \int_0^{\infty} d P'
\sum_{w \in \frac{\zi}{2} }
\cos \frac{4 \pi P P'}{k}  \nonumber \\
&&  \times \frac{\cos^2 2 r P \cosh^2 \pi P +
(-1)^{2w} \sin^2 2 r P \sinh^2 \pi P}{
\sinh 2 \pi P \sinh \frac{2 \pi P}{k}}\ ch^{NS}_c (P',kw; \tau ,\nu )  .
\end{eqnarray}
Since the transformation properties of continuous $N=2$ characters are
analogous to the transformation properties of purely bosonic coset characters,
we can discuss the results similarly as for the bosonic coset
\cite{Ribault:2003ss}.  To make the discussion more explicit,
 and in particular to match on to the boundary reflection amplitude, we follow
\cite{Ponsot:2001gt,Ribault:2003ss} to check the relative Cardy
condition. We compare the regularized
density of states obtained in the open string channel with the one we expect
from a reflection amplitude which is the natural $N=2$ generalization
of the reflection amplitude in
\cite{Ponsot:2001gt,Ribault:2002ti,Ribault:2003ss} (which
satisfies factorization).
To this end,
we subtract a reference amplitude labeled by $r^{\ast}$,
and use trigonometric identities and the change of variables
$t=2 \pi P$ to obtain:
\begin{eqnarray}
&& e^{- \frac{c i \pi \nu^2}{3 \tau} }
(Z_{r,\theta_0}^{D1}-Z_{r^{\ast},\theta_0}^{D1})
 =  \frac{8 {\cal N}_1^2}{k}
\int_0^{\infty} dP' \ \sum_{w \in \frac{\zi}{2} } \ ch^{NS}_c  (P',kw) \,\, \times\\
&& \frac{\partial}{\partial P'} \int_0^{\infty} \frac{dt}{t}
\frac{k}{16 \pi} \frac{\cosh^2 \frac{t}{2}-(-1)^{2w} \sinh^2
   \frac{t}{2}}{\sinh t \sinh \frac{t}{k}}
\left\{ \sin \frac{2t}{k}
(P'+\frac{rk}{\pi}) +  \sin \frac{2t}{k}
(P'-\frac{rk}{\pi}) - (r \rightarrow r^{\ast}) \right\}.\nonumber
\end{eqnarray}
To link the relative density of continuous states in the open string channel
to the boundary reflection amplitude it is convenient to define the special
functions:
\begin{eqnarray}
\log S_k^{(0)}(x) &=& i \int_0^{\infty}\frac{dt}{t}
\left( \frac{\sin \frac{2tx}{k}}{2 \sinh \frac{t}{k} \sinh t}-\frac{x}{t}
\right)\, ,
\nonumber \\
\log S_k^{(1)}(x) &=& i \int_0^{\infty}\frac{dt}{t}
\left( \frac{\cosh t \sin \frac{2tx}{k}}{2 \sinh \frac{t}{k} \sinh
  t}-\frac{x}{t} \right).
\end{eqnarray}
We can then write our ansatz for the boundary reflection amplitudes
as:
\begin{eqnarray}
R(P,w \in \zi |r) &=& \nu_k^{iP}
\frac{\Gamma_k^2(-\frac{1}{2}-iP+k)
 \Gamma_k(2 i P + k) S_k^{(0)}(P+
  \frac{rk}{\pi})}{\Gamma_k^2(\frac{1}{2}+iP+k) \Gamma_k(-2iP+k)
 S_k^{(0)}(-P+ \frac{rk}{\pi})}
\nonumber \\
\tilde{R} (P,w \in \zi +\frac{1}{2}|r)
&=& \nu_k^{iP} \frac{\Gamma_k^2(-\frac{1}{2}-iP+k)
 \Gamma_k(2 i P + k)
S_k^{(1)}(  P+ \frac{rk}{\pi})}{
\Gamma_k^2(\frac{1}{2}+iP+k) \Gamma_k(-2iP+k)
 S_k^{(1)}(-P+ \frac{rk}{\pi})}.
\label{reflecamplD1}
\end{eqnarray}
For the definition of the generalized gamma-functions $\Gamma_k$,
we refer to e.g. \cite{Ribault:2003ss} -- they immediately drop
out of the computation of the relative partition function. Using
the reflection amplitudes (and the fact that they are parity odd)
we can show that the relative Cardy condition holds:
\begin{eqnarray}
e^{- \frac{c i \pi \nu^2}{3 \tau} }
(Z_{r,\theta_0}^{D1}-Z_{r^{\ast},\theta_0}^{D1})
&=& \frac{{\cal N}_1^2}{\pi i} \int_0^{\infty} dP'
\left\{ \sum_{w \in  \zi \phantom{\frac{1}{2}}} \left(\frac{\partial}{\partial P'}
 \log \frac{R(P',w |r)}{R(P',w|r^{\ast})} \right) ch_{c}^{NS} (P',kw)
 \right.
 \nonumber \\
& & \left. + \sum_{w \in \zi +\frac{1}{2}}
 \left( \frac{\partial}{\partial P'}
 \log \frac{\tilde{R}(P',w|r)}{\tilde{R}(P',w|r^{\ast})}\right) ch^{NS}_c (P',kw) \right\}.
\label{ospfd1}
\end{eqnarray}
To obtain agreement with the density of states expected on the
basis of the boundary reflection amplitude, we can fix ${\cal
N}_1^2=1/2$.
Notice that in the open string channel, a
pure winding state has $J^3_{0,open}=2J^3_{0,closed}=kw$. So we get in
(\ref{ospfd1}) contributions
from open strings winding both integer and half-integer times around the cigar.
This is consistent with the semiclassical geometry of the D1 branes in the cigar
\cite{Ribault:2003ss}.

We have thus verified that the relative
Cardy condition is satisfied by our $D1$-branes. In summary, in
this section we have argued that the relative Cardy condition
holds, given the one-point functions for the D1-branes we started
out with, and the extension of the boundary reflection amplitudes
of \cite{Ponsot:2001gt,Ribault:2002ti} to $N=2$ theories. The
computation follows the lines of \cite{Ribault:2003ss} due to the
close connection between (continuous) $N=2$ characters and those
of the bosonic coset, and their modular properties (see comment at
the end of Section 4). Note that the D1-branes couple to
continuous bulk states with zero winding only.

\section{D2-branes}
\label{D2}

In this section, we analyze the Cardy condition for the D2-branes which we
can construct from the $H_2$ branes in Euclidean $AdS_3$. They correspond
to type B branes w.r.t. the N=2 superconformal algebra.
We find some puzzling features when trying to perform the Cardy check.
Following \cite{Ribault:2003ss} and the general logic outlined before,
we propose the following one-point function for the D2-branes
parameterized by $\sigma$, in the NS sector~:
\begin{multline}
\label{oneptD2}
\langle \Phi^{j}_{nw} (z,\bar{z} )\rangle^{D2}_{\sigma}  = \delta_{n,0}\
\frac{\Psi_{\sigma} (j,w)}{|z-\bar{z}|^{\Delta_{j,w}}} \ , \quad
\text{with}\ :\\
\Psi_{\sigma} (j,w)
= {\cal N}_2 \ \nu^{\frac{1}{2}-j} \,  \frac{\Gamma(j+\frac{kw}{2})\Gamma(j-\frac{kw}{2})
}{\Gamma(2j-1) \Gamma(1-\frac{1-2j}{k})}
\frac{e^{i \sigma(1-2j)} \sin \pi (j-\frac{kw}{2}) +
      e^{-i \sigma(1-2j)} \sin \pi (j+\frac{kw}{2})
}{\sin \pi(1-2j) \sin \pi \frac{1-2j}{k} }
\end{multline}
This one point function has poles corresponding to the discrete
representations, and therefore will couple both to localized and extended
states. This is expected on general grounds since these
 D2-branes carry D0-brane
charge.

The annulus partition function in the closed string channel, for general
Casimir labeled by $j$, in the NS sector, is:
\begin{subequations}
\begin{align}
Z^{D2}_{\sigma \sigma'} (-1/\tau, \nu/\tau )
&= -k {\cal N}_2^2 \ \int dj \sum_{w \in \zi}
\frac{ch^{NS} \left(j,\frac{kw}{2};-1/\tau, \nu/\tau \right)}{\sin \pi (1-2j) \sin \pi
  \frac{(1-2j)}{k}} 
\\
\hskip-2.6cm &
 \left\{ 2 \cos(\sigma+\sigma')(1-2j) -2 \cos (\sigma-\sigma') (1-2j) \cos 2 \pi j
\phantom{\frac{1}{2}}  \right. \label{l2cardD2}
\\
\hskip1.4cm
&\left.
 + \ \frac{2 \cos (\sigma -\sigma') (1-2j) \sin^2 2 \pi j}{ \cos \pi kw-\cos 2 \pi
j}
- \frac{2 i \sin(\sigma-\sigma')(1-2j) \sin 2 \pi j \sin \pi kw}{
\cos \pi kw - \cos 2 \pi j} \label{l3cardD2}
\right\}
\end{align}
\end{subequations}
We can read from this expression that the different terms will contribute in a
very different fashion.

\subsection{D1-like contribution}
The two terms of the second line~(\ref{l2cardD2}) will
give a contribution similar to the D1-branes (which is an expected contribution, on the basis
of the fact that the D2-branes also stretch along the radial direction), with imaginary parameter
though. Explicitely, we have in the closed channel~:
\begin{multline}
Z^{D2,(b)}_{\sigma \sigma'} (-1/\tau, \nu/\tau ) =
2k {\cal N}_{2}^2 \ \int_{0}^{\infty} dP \sum_{w \in \zi}
\frac{ch^{NS}_c \left(P,\frac{kw}{2};-1/\tau, \nu/\tau \right)}{\sinh 2\pi P
  \sinh 2\pi P/k} \\ \left[ \cosh 2P(\sigma + \sigma') + \cosh 2P(\sigma -
  \sigma') \cosh 2\pi P \right]\, .
 \end{multline}
As for the D1-branes, we consider the \emph{relative partition function}
w.r.t. the annulus amplitude for reference branes of parameters $(\sigma_0 , \sigma_{0}')$.
Going through the same steps as in the previous sections, we obtain the
open string channel amplitude~:
\begin{align}
Z^{D2,(b)}_{\sigma \sigma'} (\tau, \nu )
=  4k {\cal N}_{2}^2
\int dP' \, \frac{\partial}{2i\pi \, \partial P'}
\log \left\{
\frac{ R ( P' | i\frac{\sigma + \sigma'}{2} ) \,
\tilde{R} ( P' | i\frac{\sigma - \sigma'}{2})}{
R ( P'| i\frac{\sigma_{0} + \sigma_{0}'}{2} ) \,
\tilde{R} ( P'| i\frac{\sigma_{0} - \sigma_{0}'}{2} )} \right\} \sum_{n \in \zi} ch^{NS}_c (P',n;\tau,\nu)
\end{align}
in terms of reflections amplitudes similar as before,
see~(\ref{reflecamplD1}), but with
imaginary parameters $i (\sigma \pm \sigma')/2$ .
We fix the normalization constant
to ${\cal N}_{2}^2 = \frac{1}{4k}$.

\subsection{D0-like contribution}
Now we concentrate on the last two terms~(\ref{l3cardD2}) of the annulus
amplitude. As we will see this will give a contribution similar to those
of D0-branes, hence with both discrete and continuous contributions. Let's
first concentrate on the latter. Since the last term is odd in $w$, it
will cancel from the amplitude\footnote{Strictly speaking, this holds only when $\nu=0$. The same is true in the discrete sector below.}
and we are left with~:
\begin{equation}
Z^{D2,(c)}_{\sigma \sigma' ,\, cont} (-1/\tau, \nu/\tau ) =
-\frac{1}{2} \int_{0}^{\infty} dP \sum_{w \in \zi}  \frac{\cosh 2P (\sigma - \sigma') \sinh^2 2\pi
  P\  ch^{NS}_c \left(P,\frac{kw}{2};-1/\tau, \nu/\tau \right)}{(\cosh 2\pi P+
  \cos \pi kw)\sinh 2\pi P \sinh 2\pi P/k}
\end{equation}
Assuming that $\sigma - \sigma' = 2\pi m/k$, $m \in \zi$, we recognize a D0
amplitude for two branes of same parameter $m$~:
\begin{multline}
Z^{D2,(c)}_{\sigma \sigma' ,\, cont} \left( -\frac{1}{\tau}, \frac{\nu}{\tau}\right) =
- \frac{1}{2} \int_{0}^{\infty} dP \sum_{w \in \zi}  \frac{\left(2\sinh^2 (2\pi P
  m/k) +1 \right) \sinh 2\pi
  P\ }{(\cosh 2\pi P+
  \cos \pi kw) \sinh 2\pi P/k} \\
\times \quad  ch^{NS}_c \left(P,\frac{kw}{2};-1/\tau, \nu/\tau \right)\, ,
\end{multline}
up to the constant term in the bracketed expression, that will drop from the relative partition
function. The normalization $-1/2$ of this expression has to be compared
with the normalization $(-)^{w(m-m)} = 1$ of the D0 computation~(\ref{D0cont}).
\subsubsection*{Discrete representations}
We can also make the identification with a D0-like contribution as
follows. First we consider the more straightforward case $w>0$. Then
we pick the poles of the discrete representations in the domain~:
$j \in D = \left( \frac{kw}{2} - \en \right) \cap
\left] \frac{1}{2} ; \frac{k+1}{2} \right[\ $. For each pole we will
have a contribution of $2\pi$ times the residue~:
\begin{equation}
-\frac{1}{2}
\sum_{j \in D} \sum_{w > 0}
(-)^{j-\frac{kw}{2}} \frac{\cos 2\pi m \frac{2j-1}{k} \, \sin
  \pi (2j -1) - i \sin 2\pi m \frac{2j-1}{k} \, \sin \pi kw}{\sin \pi ( j +
  \frac{kw}{2}) \sin \pi \frac{2j-1}{k}}
 ch_{d}^{NS} (j,-kw/2-j)
\end{equation}
We write $j=\frac{kw}{2}-r$, with $r\in \en$. Then for every $r$,
there is only one value of $w$, that we'll call $w_r$ such that
$j$ is in the correct range. Explicitely $w_r$ is given by:
$w_r =  \lfloor \frac{2r+1}{k}\rfloor +1$. We call also $j_r$
the value of $j$ that has been picked. With this procedure we get~:
\begin{equation}
-\frac{1}{2} \sum_{r \in \mathbb{N}} (-)^{w_r}
\frac{\cos 2\pi m \frac{2r+1}{k} - i \sin  2\pi m \frac{2r+1}{k}}{\sin \pi
  \frac{2r+1}{k}}
ch_{d}^{NS} (j_r,r)
\end{equation}
Let us now consider the case $w \leqslant 0$. In this case we have to use the
isomorphism of representations~: $\mathcal{D}_{j}^{+,w} \, \simeq \,
\mathcal{D}_{j'}^{-,w-1}$, with $j' = \frac{k+2}{2}  - j$. We have then
$j' = -\frac{k}{2} (w-1) -r'$, $r'\in \mathbb{N}$,
and the value of $w$ is fixed to~: $w_{r'} = -\lfloor \frac{2r'+1}{k} \rfloor$. This leads to
the following contribution to the annulus amplitude:
\begin{equation}
-\frac{1}{2} \sum_{r' \in \mathbb{N}} (-)^{w_r'-1}
\frac{\cos 2\pi m \frac{2r'+1}{k} + i \sin  2\pi m \frac{2r'+1}{k}}{\sin \pi
  \frac{2r'+1}{k}}
ch_{d}^{NS} (j'_{r'},-r')
\end{equation}
Then it is possible to add the two contributions, which cancels the imaginary
part, and leaves us with~:
\begin{equation}
Z^{D2,(c)}_{\sigma \sigma' ,\, disc} \left( -\frac{1}{\tau},
\frac{\nu}{\tau}\right) =
- \frac{1}{2} \sum_{r \in \mathbb{\zi}} (-)^{\lfloor \frac{2r+1}{k} \rfloor}
\frac{2\sin^2 2\pi m \frac{r+1/2}{k} - 1}{\sin 2\pi
\frac{r+1/2}{k}}
ch_{d}^{NS} \left( j_r,r;-\frac{1}{\tau},\frac{\nu}{\tau} \right).
\end{equation}
This is again $-1/2$ of the expression of the amplitude for two D0's of
parameter $m$, eq.~(\ref{sumdisc2}), up to the irrelevant constant term.

\subsection*{Some comments on D2-branes physics}
The following comments are in order:
\begin{itemize}
\item in the computation we assumed that the difference of the parameters of
  the D2 in the annulus amplitude is quantized: $\sigma' - \sigma = 2\pi m
  /k$, $m \in \zi$. This relative quantization condition is discussed
  in~\cite{Ribault:2003ss} for the bosonic coset. Indeed the difference
  of the D2-branes parameters is the net induced D0-charge, hence it should be
  quantized. To be more precise it is believed that a D2-brane with parameter
  $\sigma'$ is a bound state of a brane of parameter $\sigma$ with $m$
  D0-branes (for the $\sigma' > \sigma$ case , otherwise one has to reverse
  the picture)

 \item as a corollary, the one-point functions for the D2-branes have poles
 \emph{both} of the localized type and of the bulk type
\item the computation of the annulus amplitude gives a continuous spectrum of
  open strings attached to the D2-branes, with a sensible density of states,
  and also a contribution similar to the D0 --~see the previous remark about
  the induced D0 charge~-- but with a normalization $(-1/2)$ which complicates
the task  of making sense
of  the open string spectrum, and blurs the physical picture of bound states. The only open string spectrum
  leading to a good physical picture would seem to correspond
 to two D2-branes with the same
  parameter $\sigma$.
\end{itemize}
Clearly further study of the physics of D2-branes is needed to clarify
their interpretation.

\section{Conclusions}
\label{conclusions}

We constructed D-branes in $N=2$ Liouville theory and the $SL(2,\mathbb{R})/U(1)$
super-coset conformal field theory, and checked the (relative) Cardy
condition as well as consistency with the proposed boundary reflection
amplitudes, which were proven earlier to satisfy the factorization
constraints.
The one-point functions that we constructed remarkably decouple
the poles in bulk amplitudes (notably the reflection amplitude) into
poles associated to infinite volume and the ones associated to normalizable
discrete states. It would be interesting to investigate whether the
boundary decoupling phenomenon aids in understanding more aspects of the
conjectured holographic duality in the  (doubly scaled) little string
theory.

We have shown (for the particular but extendable) case of the D0-branes
how to generalize the results to other sectors of the supersymmetric
Hilbert space, and to orbifolds of the cigar conformal field theory.

We have pointed out throughout this work some similarities of the one-point functions
for $N=2$ Liouville with those of the $N=0,1$ cases, like the decoupling of the bulk poles.
It would be interesting to see also whether the interesting relations
uncovered in \cite{Martinec:2003ka,Seiberg:2003nm} between
the boundary states associated to localized and extended branes have also
any manifestation in the $N=2$ theory.

In appendices,
we argued for the general use of the $SL(2,\mathbb{R})$ symmetry that can be obtained
by enhancing $N=2$ theories with an orthogonal free
scalar, and we suggested that this tool provides some technical support for an
attempt to interpret the variables ($x,\bar{x}$) --~parameterizing
the $SL(2,\mathbb{R})$ quantum numbers in position space on the boundary of
AdS$_3$~-- as new worldsheet variables.
We showed in great detail the fact that super-coset characters agree with
$N=2$ characters, and the $N=2$ spectral flow has a natural interpretation
in terms of $SL(2,\mathbb{R})$ quantum numbers.

Note that we have constructed D-branes for generic
values of the level $k$.
It is known that the conformal field theory correlators depend strongly on
whether $k$ is rational or irrational (for instance via the shift equations,
or the structure of the poles in the bulk correlators). It seems important
to further clarify the relation between the construction of branes at rational
and irrational values of $k$ (for instance by further comparing
the techniques developed in \cite{Israel:2004xj} to the results of
\cite{Ribault:2003ss}), as well as the dependence of bulk correlators on
this most intriguing distinguishing feature.

The branes we constructed form an integral part of the construction of
D-branes in non-compact non-trivially curved
supersymmetric string theory backgrounds. One particular
application amongst these is the construction of D1-branes and D3-branes
 in the
Little String Theory background in the double scaling limit (see also
e.g. \cite{Elitzur:zh,Pelc:2000kb,Ribault:2003sg}). Indeed,
when we consider the  doubly scaled limit for NS5-branes
which are evenly distributed over a topologically trivial circle, we
obtain a closed string background  which is supposed to represent
a holographic dual of the Higgs phase of the Little String Theory living
on the NS5-branes~\cite{Giveon:1999px}, whose partition function
has been studied in~\cite{Eguchi:2004yi,Israel:2004ir}.
The role of the W-bosons in the LST is played by the
D1-branes stretching between the NS5-branes. One can construct at least
the one-point functions corresponding to these D-branes by combining the
D0-brane in the $SL(2,\mathbb{R})/U(1)$ theory and a D1-brane in the supersymmetric
$SU(2)/U(1)$ coset theory, properly taking into account the discrete orbifold
operation. The construction of these one-point functions is now
straightforward.
 We will return to some of these issues in the near future~\cite{LSTbranes}.

\section*{Acknowledgments}
We thank Shmuel Elitzur, Angelos Fotopoulos, Stefan Fredenhagen, Gast\'on Giribet, Amit Giveon,
Anton Kapustin, Elias Kiritsis, Costas Kounnas, David Kutasov, Marios
Petropoulos, Boris Pioline, Sylvain Ribault, Adam Schwimmer and especially
Volker Schomerus for comments and discussions.
AP thanks the Laboratoire de Physique Th\'eorique
de l'\'Ecole Normale Sup\'erieure for hospitality during the first part of
this work, and the organizers of the ESI Workshop  "String Theory in Curved
Backgrounds and Boundary Conformal Field Theory". AP is supported by the Horowitz Foundation
and by an ESI Junior Fellowhip.
This work is supported in part by the Israeli Science Foundation and by the EEC under
the contracts HPRN-CT-2000-00122, HPRN-CT-2000-00131.


\appendix


\boldmath
\section{Computing N=2, c$>$3 characters }
\label{characters}
\unboldmath

In this appendix we will compute the $N=2,c>3$ characters
appearing in the spectrum of the D-branes studied in the paper.
We will obtain them as characters of
the supersymmetric coset \slc,
through a rather standard procedure.
The same characters can be computed
by subtracting the null-vectors modules from the free action of the
$N=2$ generators, as has been done in \cite{Dobrev:wd,Dobrev:1986hq,Kiritsis:1986rv}.
The $N=2$ representations that we consider are both unitary and non-unitary.
The characters of the former have been computed in
those papers and coincide with our results, while
for non-unitary representations we present the characters for the first time.
The fact that the coset yields {\it irreducible}
$N=2$ characters can be traced back to the fact that the \slr\ characters
we start with correspond to irreducible representations of the \slr\ algebra.

Notice that we have here a non-minimal
version of a similar situation for
($N=1,2$ supersymmetric) unitary minimal models,
where the (supersymmetric) Virasoro characters coincide with the characters
of certain cosets involving $SU(2)$ factors which realize the minimal models \cite{Goddard:vk}.

Moreover, this coincidence of the two ways of computing the characters,
which was independently noticed in \cite{Eguchi:2004yi} and \cite{Israel:2004ir}, is central
to the fact that the D-branes that we build in this paper belong to the class of objects,
such as the correlation functions \cite{FZZ,Giveon:2001up,Fukuda:2001jd}
or the modular-invariant partition function \cite{Eguchi:2004yi,Israel:2004ir},
for which there is no distinction
as to whether we are in the $N=2$ Liouville or in the supersymmetric cigar.

Let us review first how the $N=2$ algebra arises in the susy
\slc\ coset \cite{Kazama:1988qp}.
The supersymmetric \slr\ model at level $k$ has currents $J^{a},\psi^{b}$ ($a,b=1,2,3$), with
OPEs
\begin{eqnarray}
J^a(z) J^b(w) &\sim &  { g^{ab} k/2 \over (z-w)^2} +  {f^{ab}_{\,\,\,\,\,\,\,\,\,c}\,  J^c(w) \over z-w}\,,\nonumber\\
J^a(z) \psi^b(w) &\sim &  { f^{ab}_{\,\,\,\,\,\,\,\,\,c}  \psi^{c}(w) \over z-w} \,, \nonumber \\
\psi^a(z) \psi^b(w) &\sim& {g^{ab} \over z-w}
\end{eqnarray}
where $g^{ab} = \mathrm{diag}(+,+,-)$, $f^{123}=1$ and indices in the antisymmetric $f^{abc}$ are raised and
lowered with $g^{ab}$.

We first define
\begin{eqnarray}
j^a &=& J^a - \hat{J}^a \nonumber \\
\hat{J}^a &=& - \frac{i}{2} f^{a}_{\,\,\,\,\,bc} \psi^b \psi^c
\label{jj}
\end{eqnarray}
The currents $j^{a}$ commute with the three fermions and  generate a {\it bosonic} \slr\ model at level $k+2$.
The currents $\hat{J}^a,\psi^a$ form a supersymmetric \slr\ model at level $-2$.
The Hilbert space of the original supersymmetric \slr$_k$ theory is the direct product of the Hilbert
space of the bosonic \slr$_{k+2}$ and that of the three free fermions.

We are interested in the coset obtained by gauging the $U(1)$ symmetry generated by $J^3,\psi^3$. This coset
has an $N=2$ algebra generated by
\be
G^{\pm} &=& \sqrt{\frac{2}{k}} \ \psi^{\pm}j^{\mp} \nn \\
J^R &=& \frac{2}{k} j^3 + \frac{k+2}{k}\,\, \hat{J}^3 = \frac{2}{k}J^3 + \psi^+\psi^-
\label{kz} \\
T &=& T_{SL(2,\mathbb{R})} - T_{U(1)} \nn
\ee
where $\sqrt{2} \psi^{\pm} = \psi^1 \pm i \psi^2$ and
\be
T_{U(1)} = -\frac{1}{k} J^3J^3 +\frac12 \psi^3 \partial \psi^3 \,.
\ee
The currents (\ref{kz}) commute with $J^3, \psi^3$ and satisfy the $N=2$ superconformal algebra
\eqn{opentwo}{\eqalign{
J^R(z) J^R(w)  & \sim  \frac{c/3 }{(z-w)^2}  \,,\cr
J^R(z) G^{\pm}(w) & \sim \pm \frac{G^{\pm}(w)}{(z-w)} \,, \cr
G^{+}(z)  G^{-}(w) & \sim
 \frac{2c/3}{(z-w)^3} + \frac{2J^R(w)}{(z-w)^2} +  \frac{1}{(z-w)}\left(2T(w) + \d J^R(w) \right) \,.
}}
with the central charge
\be
c= 3 + \frac{6}{k} \,.
\ee
A highest-weight representation of the $N=2$ algebra with a given central charge $c$
is  determined by the conformal dimension $h$ and the $U(1)$ charge $Q$
of the highest weight state.
By building the $N=2$ representations through the \slc\ coset,
we can parameterize $h,Q$ by means of \slr\ quantum numbers
as follows from the zero modes of $J^3,T$ in (\ref{kz}).

The states of the  coset  are all those states in the parent \slr\ theory
annihilated by the modes $J^3_{n>0}, \psi^3_{n>0}$,
and a {\it primary} state of the coset is a coset state
which is also a primary of the $N=2$ algebra.
Every primary state of the parent theory is clearly a primary of the coset.
In addition, for the discrete and finite-dimensional representations of
 $j^a$, there are also {\it descendent} states of the parent theory which are {\it primaries} of the coset.

The Hilbert space of the parent theory can be decomposed into
subspaces with definite $J^3_0$ eigenvalue $m$.
Since the $N=2$ algebra commutes with $J^3_0$,
from a given \slr\ representation with spin $j$,
we will obtain a different $N=2$ representation
for each value of $m$. The use of $j,m$  labels is most
convenient, because performing integer spectral flow in $N=2$
just amounts to an integer shift in $m$, as we will see below.

For a given \slr\ representation, we are interested in the characters
\be
ch_{j,m}^{NS}(q,y) &=& q^{-{c \over 24}} \mathrm{Tr}_{NS} \,\,   q^{L_0} y^{J^R_0} \\
ch_{j,m}^{\nst}(q,y) &=& q^{-{c \over 24}} \mathrm{Tr}_{NS} \,\,  (-1)^F q^{L_0} y^{J^R_0} \\
ch_{j,m}^{R}(q,y) &=& q^{-{c \over 24}} \mathrm{Tr}_R \,\,   q^{L_0} y^{J^R_0}
\ee
where the trace is taken on the Hilbert space of the coset.
We will compute in all the cases  the NS character first. The Ramond
characters will be obtained by half-spectral flow \cite{Schwimmer:mf} as
\be
ch^{R}(\tau,\nu)= q^{\frac{c}{6}(\frac12)^2}y^{\frac{c}{3}\frac12}ch^{NS}(\tau, \nu + \frac{\tau}{2})
\ee
where $q=e^{i 2 \pi \tau}, y=e^{i 2 \pi \nu}$.
As for the  \nstr\ characters, it is easy to see that
they are given by
\be
ch_{j,m}^{\nst}(\tau,\nu) =  e^{i \pi Q_{j,m}} ch_{j,m}^{NS}(\tau,\nu - 1/2)
\ee
where $Q_{j,m}$ is the $U(1)$ charge of the highest weight state.
The NS characters can be obtained from
\begin{eqnarray}
\chi_j(q,x,y)&=&  \mathrm{Tr} \,\,  q^{L_0-{c + 3/2 \over 24}} \,\, x^{J^3_0}y^{J^R_0}
= \sum_{m} x^{m}\, \xi_m(q) \, ch_{j,m}^{NS}(q,y)  \label{fullchar}
\end{eqnarray}
where
\begin{eqnarray}
\xi_m(q) = q^{- \frac{3}{48} -\frac{m^2}{k}} \prod_{n=1}^{\infty} \frac{(1+q^{n-\frac12})}{(1-q^n)}
\end{eqnarray}
The trace in (\ref{fullchar}) is taken on the whole NS Hilbert space of the supersymmetric \slr$_{k}$ model
and $J^R_0$ is well-defined even before going to the coset.
We have expanded
$\chi_j(q,x,y)$ into terms with definite $J^3_0$ charge $m$ and factored
in each such term a supersymmetric $U(1)$ character $\xi_m(q)$ with highest weight $\Delta= -\frac{m^2}{k}$,
generated by the modes of $J^3, \psi^3$.

The computation of $\chi_j(q,x,y)$ itself goes as follows.
In the factorized \slr$_{k+2} \otimes \{\psi^a \}$ theory we can easily compute
\begin{equation}
\chi_j(q,z,w) = q^{-{c + 3/2 \over 24}} \,\, \mathrm{Tr} \,\,  q^{L_0}
z^{j^3_0} \times \mathrm{Tr} \,\,
q^{L_0} w^{\hat{J}^3_0}\,,
\label{eqfac}
\end{equation}
and then from (\ref{jj}) and (\ref{kz}) it follows that $\chi_j(q,x,y)$ is
given from $\chi_j(q,z,w)$ by the replacements
\begin{eqnarray}
z & \rightarrow xy^{\frac{2}{k}} \,, \nonumber \\
w & \rightarrow xy^{\frac{k+2}{k}} \,. \label{replac}
\end{eqnarray}
In (\ref{eqfac}), the trace 
over the fermions NS Hilbert space is
\be
\mathrm{Tr} \,\,  q^{L_0} w^{\hat{J}^3_0} &=&
\prod_{n=1}^{\infty} (1+q^{n-\frac12}) (1+wq^{n-\frac12})(1+w^{-1}q^{n-\frac12}) \nn \\
&=& \prod_{n=1}^{\infty} \frac{(1+q^{n-\frac12})}{(1-q^n)} \sum_{p \in \mathbb{Z}} w^p q^{\frac{p^2}{2}}\,.
\ee
For the first factor in (\ref{eqfac}), we should consider the different
possible representations of the $j^a$ algebra.
We take $q=e^{i 2 \pi \tau}, y=e^{i 2 \pi \nu}$.

\ni
\subsection*{Continuous representations}
The $j^a$ representations are built by acting with $j^a_{n<0}$ on
representations of the zero modes~$j^a_0$. For the continuous representations
we have $j=\frac12 + iP$, $P \in \mathbb{R}^+$. The \slr\ primaries are
$|P,m \rangle$, with
$m=r + \alpha, \,\, r \in \mathbb{Z}, \, \alpha \in [0,1)$
and conformal weight
\be
-\frac{j(j-1)}{k} = \frac{\frac14 + P^2 }{k} \,.
\ee
These primaries of $j^a$ are multiplied by the fermionic NS vacuum to get the full primaries,
so $j^3_0= J^3_0=m$ on these states.
Considered as primaries of $N=2$, they give rise to unitary $N=2$
representations \cite{Dixon:1989cg} with (see (\ref{kz}))
\begin{equation}
h_{P,m} = \frac{\frac14 + P^2 + m^2}{k} \,, \qquad Q_{m}= \frac{2m}{k}
\end{equation}
The character of the susy \slr\ representation is
\be
\chi_j(q,z,w) &=& q^{-{c + 3/2 \over 24}} \,\, \mathrm{Tr} \,\,  q^{L_0}
z^{j^3_0} \times \mathrm{Tr} \,\,  q^{L_0} w^{\hat{J}^3_0}\,, \nn \\
&=& q^{-{c + 3/2 \over 24}} q^{\frac{1/4 + P^2}{k}} \sum_{r,p \in \mathbb{Z}}
z^{\alpha+ r} w^{p} q^{\frac{p^2}{2}}
\prod_{n=1}^{\infty} \frac{(1+q^{n-\frac12})}{(1-q^n)^4}
\ee
and applying the steps described above it is immediate to obtain the $N=2$ characters
\begin{eqnarray}
\label{carcon1}
ch_{c}^{NS} (P,m ; \tau ,\nu ) &=& \mathrm{Tr}\,\, q^{L_0-c/24}\, y^{J_0^R}
\nonumber \\
&=& q^{\frac{P^2 + m^2}{k}} y^{\frac{2m}{k}} \frac{\vartheta_3(\tau,\nu )}{\eta(\tau)^3} \\
ch_{c}^{\nst} (P,m ; \tau ,\nu )&=& q^{\frac{P^2 + m^2}{k}} y^{\frac{2m}{k}}
\frac{\vartheta_4(\tau,\nu )}{\eta(\tau)^3} \\
ch_{c}^{R} (P,m' ; \tau ,\nu )&=& q^{\frac{P^2 + m'^2}{k}} y^{\frac{2m'}{k}}
\frac{\vartheta_2(\tau,\nu )}{\eta(\tau)^3}
\end{eqnarray}
where $m'=m + \frac12$ when the Ramond character is obtained by half-spectral flow from~(\ref{carcon1}).

\vskip 0.5cm

\ni
\subsection*{Discrete representations}
Let us consider discrete lowest-weight representations ${\cal D}^+_j$ of $j^a$,
with primaries $|j,r \rangle$, with $m=j + r$ and $r \in \mathbb{Z}, r>0$, and
with  $j \in \mathbb{R}, j>0$.
For $0 <j < \frac{k+2}{2}$ they
give rise to $N=2$  unitary representations \cite{Boucher:1986bh,Dixon:1989cg,Pakman:2003cu}.
This bound is further constrained in physical settings to be $\frac12 <j < \frac{k+1}{2}$.
For each value of $m=j+r, r \in \mathbb{Z}$ the $N=2$ primaries are
\be
\begin{array}{lll}
r \geq 0 &\qquad&  |j,j+r \rangle \\
r < 0 &\qquad&  (j^-_{-1})^{-r-1}\psi^-_{-\frac12}|j,j \rangle
\label{rangesdiscrete}
\end{array}
\ee
The quantum numbers are:
\begin{eqnarray}
\label{primdis}
\begin{array}{lrclrcl}
r \geq 0 & \qquad h_{j,r} &=& \frac{-j(j-1) + (j+r)^2}{k}
\qquad \qquad \qquad &  \qquad Q_{j,r} &=& \frac{2(j+r)}{k}  \\
r < 0 & \qquad h_{j,r} &=& \frac{-j(j-1) + (j+r)^2}{k} -r - \frac12
\qquad & Q_{j,r}&=& \frac{2(j+r)}{k} -1  \\
& &=& \frac{-(\frac{k+2}{2} -j)(\frac{k+2}{2}-j-1)+(\frac{k+2}{2}-j-r-1)^2}{k}
& &=& -\frac{2(\frac{k+2}{2} -j -r -1)}{k}
\end{array}
\end{eqnarray}
The second line in the $r<0$ case shows that these states are similar to the $r \geq 0$ states when built from
${\cal D}^-_{\frac{k+2}{2} -j}$.
Note that $r=0,-1$ correspond to
chiral and anti-chiral primaries respectively, as follows from $h_{j,0}=Q_{j,0}/2$ and
$h_{j,-1}=-Q_{j,-1}/2$.
These states are mapped to fermionic null states of relative charge $\pm 1$ along the spectral flow orbit.

To compute the characters we start with
\begin{eqnarray}
\chi_j(q,z,w) &=& q^{-{c \over 24}} \,\, \mathrm{Tr} \,\,  q^{L_0} z^{j^3_0}
  \times
\mathrm{Tr} \,\,  q^{L_0} w^{\hat{J}^3_0} \nonumber \\
&=&  { q^{-{c \over 24} -\frac{j(j-1)}{k}} z^j  \over
\prod_{n=1}^{\infty}(1-zq^{n-1})(1-z^{-1}q^{n})(1-q^n)}
\times \mathrm{Tr} \,\,  q^{L_0} w^{\hat{J}^3_0}
\label{chardis}
\end{eqnarray}
Using\footnote{See \cite{Pakman:2003kh} for a proof of this identity.}
\begin{equation}
{1 \over {\prod}_{n=1}^{\infty} ( 1-q^{n-1}z)(1-q^nz^{-1} )} = {1
\over \prod^{\infty}_{n=1}(1-q^n)^2} \sum^{\infty}_{t=-\infty}  z^{t} S_t(q) \,,
\end{equation}
where
\begin{equation}
S_t(q) = \sum^{\infty}_{s=0} (-1)^s q^{\frac12 s(s+2t+1)}
\end{equation}
the character (\ref{chardis}) can be expanded into
\begin{equation}
\chi_j(q,z,w) = q^{-\frac{c}{24} - \frac{j(j-1)}{k}} \prod_{n=1}^{\infty}
(1+q^{n-\frac12})
\sum_{p,t \in \mathbb{Z}} \frac{S_t(q) \, q^{p^2 \over 2} \,
z^{j+t} \, w^{p} }{\prod_{n=1}^{\infty} (1-q^{n})^4}
\end{equation}

\ni
After the replacement (\ref{replac}), and  defining $r=t+p$, we get the
decomposition as in (\ref{fullchar})
\be
\chi_j(q,x,y) = \sum_{r \in \mathbb{Z}} x^{j+r}\, \xi_{j+r}(q) \, \times
 \frac{q^{  \frac{-(j-1/2)^2 + (j+r)^2}{k} } y^{\frac{2(j+r)}{k} }} {\eta(\tau)^3}
\sum_{p \in \mathbb{Z}}  S_{r-p}(q) y^p q^{\frac{p^2}{2}} \,\,.
\ee
\ni
Using
\be
\sum_{p \in \mathbb{Z}}  S_{r-p}(q)\, y^p q^{\frac{p^2}{2}} = \frac{
\sum_{p \in \mathbb{Z}} y^p q^{\frac{p^2}{2}}  }{1+yq^{r+1/2}}
= \frac{\vartheta_3(\tau,\nu)}{1+yq^{r+1/2}}
\ee
we get finally
\begin{eqnarray}
ch_{d}^{NS} (j,r ; \tau ,\nu )
&=& \mathrm{Tr}\,\, q^{L_0-c/24}y^{J_0^R} \nonumber \\
&=& q^{- \frac{(j-1/2)^2}{k} + \frac{(j+r)^2}{k}} y^{2(j+r)\over k}
\frac{1}{1+yq^{1/2+r}} \frac{\vartheta_3(\tau,\nu )}{\eta(\tau)^3} \,,
\label{cardis1}
 \\
&=& q^{- \frac{(j-1/2)^2}{k} + \frac{(j+r)^2}{k}-r-\frac12} y^{{2(j+r)\over k}-1}
\frac{1}{1+y^{-1} q^{-1/2-r}} \frac{\vartheta_3(\tau,\nu )}{\eta(\tau)^3} \,. \nn
\end{eqnarray}
We wrote the characters in two forms, each form reflecting the
structure of the representation for a different range of $r$ (see (\ref{rangesdiscrete})).
For the other sectors we get
\be
ch_{d}^{\nst} (j,r ; \tau ,\nu ) &=& q^{- \frac{(j-1/2)^2}{k} + \frac{(j+r)^2}{k}} y^{2(j+r)\over k}
\frac{1}{1-yq^{1/2+r}} \frac{\vartheta_4(\tau,\nu )}{\eta(\tau)^3} \\
ch_{d}^{R} (j,r' ; \tau ,\nu ) &=& q^{- \frac{(j-1/2)^2}{k} + \frac{(j+r')^2}{k}} y^{2(j+r')\over k}
\frac{1}{1+yq^{1/2+r'}} \frac{\vartheta_2(\tau,\nu )}{\eta(\tau)^3}
\ee
where $r'=r+\frac12$ when the Ramond character is obtained by half-spectral flow from (\ref{cardis1}).

\vskip 0.5cm

\subsection*{Finite dimensional representations}
In this case the spin takes the values
$j=-\frac{(u-1)}{2}, u= 1,2...$. The highest weights are $u$-dimensional representations of
 $j_0^a$  given by $|j,m\rangle$, with
$m= j, j+1, \cdots -j$. Only for $u=1$ the induced  $N=2$ representation is unitary.
For every $m=j+r, r \in \mathbb{Z}$ the $N=2$ primaries are
\be
\begin{array}{ll}
r<0 \qquad \qquad & (j^-_{-1})^{-r-1}\psi^-_{-\frac12}|j,j \rangle  \\
0 \leq r \leq u-1 & |j, j+r \rangle  \\
r > u-1 & (j^+_{-1})^{r-u}\psi^+_{-\frac12}|j,-j \rangle
\end{array}
\label{fdntw}
\ee
with quantum numbers
\be
\begin{array}{llll}
r < 0 &&  h_{j,r} = \frac{-j(j-1) + (r+j)^2}{k} -r - \frac12 & Q_{j,r}= \frac{2(r +j)}{k} -1 \\
0 \leq r \leq u-1  &&  h_{j,r} = \frac{-j(j-1) + (r+j)^2}{k} & Q_{j,r}= \frac{2(r +j)}{k}  \\
r > u-1 &&  h_{j,r} = \frac{-j(j-1) + (r+j)^2}{k} +r -u + \frac12 \qquad  & Q_{j,r}= \frac{2(r +j)}{k} +1
\end{array}
\ee
In order to compute the characters, we start with
\begin{eqnarray}
\label{charfin}
\chi_j(q,z,w) &=& q^{-{c \over 24}} \,\, \mathrm{Tr} \,\,  q^{L_0} z^{j^3_0}
\times
\mathrm{Tr} \,\,  q^{L_0} w^{\hat{J}^3_0}  \\
  =&&  \! \! \! \frac{ q^{-{c \over 24} -\frac{j(j-1)}{k}}
    (z^{-\frac{(u-1)}{2}} -
z^{\frac{u+1}{2}}) }{\prod_{n=1}^{\infty}(1-zq^{n-1})(1-z^{-1}q^{n})(1-q^n)}
\times \prod_{n=1}^{\infty} \frac{(1+q^{n-\frac12})}{(1-q^n)}
\sum_{p \in \mathbb{Z}} w^p q^{\frac{p^2}{2}} \nn
\end{eqnarray}
The expansion goes along similar lines to the discrete representations and is left as an exercise.
The resulting $N=2$ characters  are
\be
\nn
\begin{array}{ccll}
ch_f^{NS} (u,r ; \tau ,\nu ) &=& \mathrm{Tr}\,\, q^{L_0-c/24}z^{J_0^R}  \\
 &=& q^{- \frac{(j-1/2)^2}{k} + \frac{(r+j)^2}{k} -r - 1/2} y^{{2(r+j)\over k}-1} &
\frac{(1-q^{u})}{(1+y^{-1}q^{-1/2-r})(1+y^{-1}q^{u-1/2-r})} \frac{\vartheta_3(\tau,\nu )}{\eta(\tau)^3} \,,
 \\
&=& q^{- \frac{(j-1/2)^2}{k} + \frac{(r+j)^2}{k}} y^{{2(r+j)\over k}} &
\frac{(1-q^{u})}{(1+y q^{+1/2+r})(1+y^{-1}q^{u-1/2-r})} \frac{\vartheta_3(\tau,\nu )}{\eta(\tau)^3} \,,
\\
&=& q^{- \frac{(j-1/2)^2}{k} + \frac{(r+j)^2}{k} +r +2j -\frac12} y^{{2(r+j)\over k}+1} &
\frac{(1-q^{u})}{(1+yq^{1/2+r})(1+y q^{r-u+1/2})} \frac{\vartheta_3(\tau,\nu )}{\eta(\tau)^3} \,.  \\
\end{array}
\\ \label{finchar}
\ee
Again, we expressed the characters
in forms reflecting the
structure of the representation for  different ranges of $r$.
These characters can also be expressed as
\be
ch_f^{NS} (u,r ; \tau ,\nu )
&=& ch_d^{NS} (j,r ; \tau ,\nu ) - ch_d^{NS} (-j+1,r -u ; \tau ,\nu ) \nn \\
&=&\frac{\vartheta_3(\tau,\nu )}{\eta(\tau)^3}
q^{\frac{s^2-su}{k}}y^{\frac{2s-u}{k}}
\left[\frac{1}{1+ yq^s}
-\frac{1}{1+ yq^{s-u}} \right]
\label{finchar2}
\ee
with $s=r + \frac12$. For the other sectors we get
\be
\nn
\begin{array}{ccll}
ch_f^{\nst} (u,r ; \tau ,\nu ) &=& q^{- \frac{(j-1/2)^2}{k} + \frac{(r+j)^2}{k}} y^{{2(r+j)\over k}} &
\frac{(1-q^{u})}{(1-y q^{+1/2+r})(1-y^{-1}q^{u-1/2-r})} \frac{\vartheta_4(\tau,\nu )}{\eta(\tau)^3}
 \\
ch_f^{R} (u,r' ; \tau ,\nu ) &=& q^{- \frac{(j-1/2)^2}{k} + \frac{(r'+j)^2}{k}} y^{{2(r'+j)\over k}} &
\frac{(1-q^{u})}{(1+y q^{+1/2+r'})(1+y^{-1}q^{u-1/2-r'})} \frac{\vartheta_2(\tau,\nu )}{\eta(\tau)^3}
\end{array}
\\ \label{otrosfinchar}
\ee
where $r'=r+\frac12$ when the Ramond character is obtained by half-spectral flow from (\ref{finchar}).

\subsection*{Spectral flow}
The $N=2$ algebra has the spectral flow automorphism \cite{Schwimmer:mf}
\be
L_n &\rightarrow& L_n + w J_n + \frac{c}{6}w^2 \delta_{n,0} \nn \\
J_n &\rightarrow &  J_n + \frac{c}{3} \delta_{n,o} \nn \\
G^{\pm}_{m} &\rightarrow & G^{\pm}_{m \pm w}
\ee
which maps
a representation into another one. The spectrum
of the flowed representations is obtained by measuring the flowed $L_0, J_0$
on the original representation.
The character of the spectrally flowed representation is then given by
\begin{eqnarray}
ch (\tau ,\nu ) \rightarrow q^{\frac{c}{6} w^2} &y^{\frac{c}{3} w}& ch (\tau ,\nu+w\tau )    \label{flow1}
\end{eqnarray}
Using (\ref{flowtheta}),
it is immediate to verify that in the characters considered above the spectral
flow by $w$ integer units is equivalent to a shift $m \rightarrow m +w$, thus
verifying our claim that  $m$ keeps track of the $N=2$ spectral flow orbit.

Finally, notice  that in the discrete and finite cases, \slr\ representations in which
the generators act freely give rise to an $N=2$ representations with
null descendents due to the semi-infinite/finite base in \slr.
To get some insight into the form of the characters, note that
the free action of the modes $L_{-n}, J_{-n}, G^{\pm}_{-n+1/2}$ on the
highest weight state gives the contribution (in the NS sector),
\begin{eqnarray}
\prod_{n=1}^{\infty}\frac{(1+q^{n-\frac12}y)(1+q^{n-\frac12}y^{-1})}{(1-q^n)^2}
= q^{1/8} \frac{\vartheta_3(\tau,\nu )}{\eta(\tau)^3}
\end{eqnarray}
In the discrete and finite representations the character is further modded out by the null descendents.


\boldmath
\section{Changing basis in $SL(2,\mathbb{R})$}
\unboldmath
In this appendix we recall how to Fourier transform the one-point functions
of \cite{Ponsot:2001gt} into the form in which they were used in
\cite{Ribault:2003ss},
and in the bulk of our paper, in our conventions.
\label{FT}
\subsection*{Reflection amplitude}
We use a quadratic Casimir for $SL(2,R)$ representation of the form:
\begin{eqnarray}
c_2=-j(j-1)
\end{eqnarray}
and we adopt conventions in which the bulk two-point functions are:
\begin{eqnarray}
\langle \Phi^j(x^1,\bar{x}^1) \Phi^{j'}(x^2,\bar{x}^2) \rangle
&=& |z_{12}|^{-4 \Delta_j} \left( \delta^2(x^1-x^2) \delta(j+j'-1)
+ \frac{B(j)}{|x_{12}|^{4j}} \delta(j-j') \right) \nonumber \\
B(j) &=& \frac{k}{\pi} \nu^{1-2j}
\frac{\Gamma(1-\frac{2j-1}{k})}{\Gamma(\frac{2j-1}{k})}
\end{eqnarray}
which leads after Fourier transformation
\begin{eqnarray}
\Phi^j_{nw} &=& \frac{1}{4 \pi^2} \int d^2 x^{j-1+m} {\bar{x}}^{j-1+\bar{m}}
\Phi^j(x, \bar{x})
\end{eqnarray}
to
\begin{eqnarray}
\langle \Phi^j_{nw} \Phi^{j'}_{n'w'} \rangle
&=& |z_{12}|^{-4 \Delta_j} \delta_{n+n'} \delta_{w+w'}
\left(  \delta(j+j'-1) + R(j,n,w) \delta(j-j')
 \right)
\nonumber \\
R(j,n,w) &=&  \nu^{1-2j} \frac{\Gamma(-2j+1)\Gamma(j+m)\Gamma(j-\bar{m})
\Gamma(1+\frac{1-2j}{k})}{\Gamma(2j-1)\Gamma(-j+1+m)\Gamma(-j+1-\bar{m})
\Gamma(1-\frac{1-2j}{k})}.
\end{eqnarray}
In the super-coset, we identify the elliptic eigenvalues of $SL(2,R)$ with
the geometric angular momentum and winding by:
\begin{eqnarray}
m &=& (n+kw)/2 \nonumber \\
\bar{m} &=& (-n+kw)/2.
\end{eqnarray}
\subsection*{Localized branes}
Starting from the one-point function for spherical branes in $H_3$ \cite{Ponsot:2001gt}:
\begin{eqnarray}
\langle \Phi^j(x|z) \rangle_s  &=& -\frac{1}{2 \pi} (1+x \bar{x})^{-2j}
\frac{\Gamma(1+ (-2j+1)/k)}{\Gamma(1-1/k)}
\frac{\sin s(-2j+1)}{\sin s} \nu^{-j+1} |z-z'|^{-2\Delta_j}, \nn \\
\end{eqnarray}
we find its Fourier transform:
\begin{eqnarray}
\langle \Phi^j_{np}(z) \rangle_s  &=& \int d^2x e^{in arg(x)}
|x|^{2j-2-ip} \langle \Phi^j(x|z) \rangle_s \nonumber \\
&=& \frac{1}{2} \delta_{n,0}
\frac{\Gamma(j-ip/2)\Gamma(j+ip/2) \Gamma(1+1/k)}{\Gamma(2j-1) \Gamma(1-(-2j+1)/k)}
\frac{\sin \pi/k  \sin s(-2j+1)}{\sin s \sin \pi (-2j+1)/k},
\end{eqnarray}
where we made use of the integral:
\begin{eqnarray}
\int_0^\infty dr r^{2j-1-ip} (1+r^2)^{-2j}
&=& \frac{1}{2} \frac{\Gamma(j-ip/2)\Gamma(j+ip/2)}{\Gamma(2j)}.
\end{eqnarray}
Up to normalization, this corresponds to the one-point function in the bosonic
and
supersymmetric coset.
\boldmath
\subsection*{Extended $AdS_2$ branes}
\unboldmath
Starting from the one-point function for extended branes in $H_3$
(where $\sigma=\mbox{sgn}(x+\bar{x})$):
\begin{eqnarray}
\langle \Phi^j(x|z) \rangle_r  &=& \frac{{\cal N} k}{\pi} |x-\bar{x}|^{-2j}
\nu^{-j+1/2}
\Gamma(1+ (-2j+1)/k)
e^{(1-2j)r\sigma} |z-z'|^{- 2 \Delta_j},
\end{eqnarray}
we find the transformed coset one-point function:
\begin{eqnarray}
\langle \Phi^j_{np}(z) \rangle_r  &=& \int dx^2 e^{in arg(x)}
|x|^{2j-2-ip} \langle \Phi^j(x|z) \rangle_r \nonumber \\
&=& 2 \pi k\delta(p) {\cal N}
\nu^{-j+1/2} \frac{\Gamma(-2j+1)\Gamma(1+(-2j+1)/k)}{\Gamma(-j+1+n/2)
  \Gamma(-j+1-n/2)}
((-1)^n e^{r (-2j+1)} +  e^{-r(-2j+1)}). \nn \\
\end{eqnarray}
We used the integrals:
\begin{eqnarray}
\int_{-\pi/2}^{\pi/2} d \phi e^{in \phi} (2 \cos \phi)^{-2j}
&=&  \frac{ \pi \, \Gamma(-2j+1)}{\Gamma(1-j+n/2)\Gamma(1-j-n/2)}
\nonumber \\
\int_0^{\infty} dr r^{-1-ip} &=& 2 \pi \delta(p).
\end{eqnarray}
Thus we have reviewed the connection between the one-point functions in
\cite{Ponsot:2001gt} and the one-point functions for the coset
\cite{Ribault:2003ss} in our conventions.

\boldmath
\section{Cardy condition for D0 branes in R/\nstr\ sectors}
\unboldmath
\label{ccrn}
In this appendix we will provide some more details of the Cardy computation for D0
branes in
 the $R$ and $\nst$ sectors, along the lines of section \ref{D0NS}.

Let us start Ramond sector in the open string channel.
For the partition function in the $(u,1)$ case we take
\be
Z^{R}_{u,1}(\tau,\nu) &=& \sum_{r \in \mathbb{Z}+ \frac12} ch_f^{R} (u,r ; \tau ,\nu ) \nn \\
&=& \frac{\vartheta_2(\tau,\nu)}{\eta(\tau)^3} \sum_{s \in \mathbb{Z}}
\frac{1}{1+yq^s} (q^{\frac{s^2 -su}{k}} y^{\frac{2s-u}{k}} - q^{\frac{s^2 +su}{k}} y^{\frac{2s+u}{k}})\,,
\label{pfsbR}
\ee
where we have summed over the whole spectral flow orbit of
the $N=2$ Ramond  characters associated with the $u$-dimensional
representation of \slr.
For the general $(u,u')$ case we make a sum as in (\ref{uut}).
The computation is very similar
to the NS/NS case, so we will only indicate the major steps.

We start with the modular transform of
$Z^{R}_{u,1}(\tau,\nu)$, given by
\begin{eqnarray}
e^{-i \pi \frac{c}{3} \frac{ \nu^2}{\tau} }  && Z_{u,1}^{R}(-\frac{1}{\tau},\frac{\nu}{\tau}) =
\frac{\vartheta_4(\tau,\nu)}{\eta(\tau)^3} \,\,\, \times \nn \\
 &&
 \!\!\!\!\!\!\!\!\!\! \frac{1}{2 i \pi}  \left[ \int_{\mathcal{C}_{-\epsilon}}
+ \int_{\mathcal{C}_{+\epsilon}} \right] dZ \
(-i \pi) \, e^{ \pi Z + \frac{2i\pi \tau }{k} Z^2}\,\,  \frac{ \sinh(2 \pi  \frac{Z}{k}u)}
{\cosh (\pi Z)  }  \frac{e^{i\pi  (i\tau Z - \nu + \frac12  )} }{\cos \pi (i\tau Z-\nu + \frac12)}
\label{contourintR}
\end{eqnarray}
The contour of integration is the same as in section \ref{D0NS}, and the integrand has poles at
$\nu -i\tau Z   =s \in \mathbb{Z} $. Expanding the last factor of the
integrand as a sum over $w$ as in (\ref{firstexp}-\ref{secexp}), taking $\varepsilon \rightarrow 0$
and shifting the contour by $\frac{iwk}{2} -i\nu_2$ in each term, we arrive, for general $(u,u')$, to a
decomposition
\begin{eqnarray}
e^{-i \pi \frac{c}{3} \frac{ \nu^2}{\tau} }
Z_{u,u'}^{R}(-\frac{1}{\tau},\frac{\nu}{\tau}) =
Z_{u,u'}^{R,c} +
Z_{u,u'}^{R,d}
\label{BJDIVR}
\end{eqnarray}
where the first term is an integral over the shifted contour and the
second is the sum of poles picked during the shift. These terms are
\begin{eqnarray}
Z_{u,u'}^{R,c} &=& \sum_{ w \in \mathbb{Z} }
 \int_{0}^{+\infty} \!\!\! \! \!\!\!\! dP
\frac{   2 \sinh(2\pi P) \sinh(2 \pi  \frac{P}{k} u ) \sinh(2 \pi  \frac{P}{k} u' ) }
{(-1)^{w(u+u'+1)} [\cosh(2 \pi P) \!+\! \cos(\pi k w)] \sinh(2 \pi  \frac{P}{k} ) }
\nn \\
&& \qquad \qquad \qquad \qquad \qquad \qquad \qquad \qquad\qquad  \times \,\,\,
q^{\frac{P^2 + (kw/2)^2}{k}} y^{\frac{2(kw/2)}{k}} \frac{\vartheta_4(\tau,\nu )}{\eta(\tau)^3}  \nn \\
 &=& \sum_{ w \in \mathbb{Z} }
 \int_{0}^{+\infty} \!\!\! dP \
 \Psi_{u}^{\nst}\left(\frac12 -iP,w \right)\Psi_{u'}^{\nst}\left(\frac12 +iP,w \right)
\  ch_c^{\widetilde{NS}} (P,\frac{wk}{2} ; \tau,\nu)
\label{D0contR}
\end{eqnarray}
and
\begin{eqnarray}
Z_{u,u'}^{R,d} &=&  \sum_{r \in \mathbb{Z} }
\frac{2 \sin \left( \frac{2 \pi}{k} (2j_r -1)u\right) \sin \left( \frac{2 \pi}{k} (2j_r -1)u'\right)}
{(-1)^{w_r(u+u'+1)} \sin \left( \frac{2 \pi}{k} (2j_r -1)\right)}
\ \   \frac{\vartheta_4(\tau,\nu )}{\eta(\tau)^3}  \frac{ y^{w_r} q^{sw_r-\frac{s^2}{k}}  }{1-yq^s} \nn \\
&=& 2 \pi \sum_{r \in \mathbb{Z} } Res \left[ \Psi_u^{\nst}(-j_r +1,w_r)\Psi_{u'}^{\nst}(j_r ,w_r) \right]
\ ch_d^{\widetilde{NS}} (j_r,r ;  \tau ,\nu )
\label{sumdisc2R}
\end{eqnarray}
where $j_r$ and $w_r$ are defined in the same way as in section \ref{D0NS} and
$s=r+\frac12$ as usual. We see thus that the Cardy condition is verified.

For the open string partition function in the $\nst$ sector, which
computes the open string Witten index \cite{Douglas:1999hq}, we start with
\be
Z^{\nst}_{u,1}(\tau,\nu) &=& \sum_{r \in \mathbb{Z}} ch_f^{\nst} (u,r ; \tau ,\nu ) \nn \\
&=& \frac{\vartheta_4(\tau,\nu)}{\eta(\tau)^3} \sum_{s \in \mathbb{Z}+ \frac12 }
\frac{1}{1-yq^s} (q^{\frac{s^2 -su}{k}} y^{\frac{2s-u}{k}} - q^{\frac{s^2 +su}{k}} y^{\frac{2s+u}{k}})
\label{pfsbNST}
\ee
Its modular transform is
\be
e^{-i \pi \frac{c}{3} \frac{ \nu^2}{\tau} }  && Z_{u,1}^{\nst}(-\frac{1}{\tau},\frac{\nu}{\tau}) =
\frac{\vartheta_2(\tau,\nu)}{\eta(\tau)^3} \nn \\
\times &&
 \frac{1}{2 i \pi}  \left[ \int_{\mathcal{C}_{-\epsilon}}
+ \int_{\mathcal{C}_{+\epsilon}} \right] dZ \
(-i \pi) \, e^{ \pi Z + \frac{2i\pi \tau }{k} Z^2}\,\,  \frac{ \sinh(2 \pi  \frac{Z}{k}u)}
{\sinh (\pi Z)  }  \frac{e^{i\pi  (i\tau Z - \nu  )} }{\cos \pi (i\tau Z-\nu )}
\label{contourintNST}
\ee
with the same contour as before, but now
 the integrand has poles at
$\nu -i\tau Z   =s \in \mathbb{Z} + \frac12$. Expanding again the last factor of the
integrand  as in (\ref{firstexp}-\ref{secexp}), taking $\varepsilon \rightarrow 0$
and shifting the contour by $\frac{iwk}{2} -i\nu_2$ in each term, we arrive, for general $(u,u')$, to
\begin{eqnarray}
e^{-i \pi \frac{c}{3} \frac{ \nu^2}{\tau} }
Z_{u,u'}^{\nst}(-\frac{1}{\tau},\frac{\nu}{\tau}) =
Z_{u,u'}^{\nst,c} +
Z_{u,u'}^{\nst,d}
\label{BJDIVNST}
\end{eqnarray}
where
\begin{eqnarray}
Z_{u,u'}^{\nst,c} &=& \sum_{ w \in \mathbb{Z} }
 \int_{0}^{+\infty} \!\!\! \! \!\!\!\! dP
\frac{   2 \sinh(2\pi P) \sinh(2 \pi  \frac{P}{k} u ) \sinh(2 \pi  \frac{P}{k} u' ) }
{(-1)^{w(u+u')} [\cosh(2 \pi P) \!-\! \cos(\pi k w)] \sinh(2 \pi  \frac{P}{k} ) }
\nn \\
&& \qquad \qquad \qquad \qquad \qquad \qquad \qquad \qquad\qquad  \times \,\,\,
q^{\frac{P^2 + (kw/2)^2}{k}} y^{\frac{2(kw/2)}{k}} \frac{\vartheta_2(\tau,\nu )}{\eta(\tau)^3}  \nn \\
 &=& \sum_{ w \in \mathbb{Z} }
 \int_{0}^{+\infty} \!\!\! dP \
 \Psi_{u}^{R^{-}}\left(\frac12 -iP,-w \right)\Psi_{u'}^{R^{+}}\left(\frac12 +iP,w \right)
\  ch_c^{R} (P,\frac{wk}{2} ; \tau,\nu)
\label{D0contNST}
\end{eqnarray}
and
\begin{eqnarray}
Z_{u,u'}^{\nst,d} &=&  \sum_{r \in \mathbb{Z} + \frac12}
(-1)^{w_r(u+u')}
\frac{2 \sin \left( \frac{ \pi}{k} (2j_r -1)u\right) \sin \left( \frac{ \pi}{k} (2j_r -1)u'\right)}
{\sin \left( \frac{ \pi}{k} (2j_r -1)\right)}
 \frac{y^{w_s}
q^{sw_s-\frac{s^2}{k}}}{1 + yq^s}
\frac{\vartheta_2(\tau,\nu)}{\eta(\tau)^3} \nn \\
 &=& 2 \pi \sum_{r \in \mathbb{Z} + \frac12} Res \left[ \Psi_u^{R^-}(-j_r +1,-w_r)\Psi_{u'}^{R^{+}}(j_r ,w_r) \right]
\ ch_d^{R} (j_r,r ;  \tau ,\nu ) \,,
\end{eqnarray}
with $j_r$ and $w_r$ defined as in section \ref{D0NS}, but now taking $r \in \mathbb{Z} + \frac12$.

In both $Z_{u,u'}^{R,d}$ and $Z_{u,u'}^{\nst,d}$ the residues are
computed when considering the bracketed expressions as analytical functions of~$j$.

\boldmath
\section{Embedding $N=2$ into \slr}
\label{embedding}
\unboldmath

Our goal in this appendix is to elaborate on the relation
between the $N=2, c>3$ chiral algebra and the \slr\ algebra,
showing how the former always lead to the latter.
These ideas go back to \cite{Dixon:1989cg}.

Consider an $N=2$ algebra with $c>3$. We can write the supercurrents and
the $U(1)$ R-current in the following form:
\eqn{cucp}{\eqalign{
G^{\pm}  & =  \sqrt{2c\over3}\pi^{\pm}e^{\pm i \sqrt{3\over c}\phi} \cr
J^R & = i \sqrt{\frac{c}3} \d \phi
}}
where we made use of a canonically normalized scalar $\phi$ and the fact
that the supercurrents carry $U(1)$ R-charge $ \pm 1$. The fields $\pi^{\pm}$
are the first parafermionic currents of \slr $/ U(1)$.
We now observe that if we add a trivial auxiliary $U(1)$ to the
theory, parameterized by
an anti-hermitean scalar field $T$,
we can define the following currents:
\be
j^3 & =& i\sqrt{k+2 \over k} \d \phi + {(k+2) \over \sqrt{2k}} \d T =
J^R + {(k+2) \over \sqrt{2k}} \d T \cr
j^{\pm} & =& \sqrt{k+2} \, \pi^{\pm}
e^{\mp \sqrt{\frac{2}{k+2}}\left( i \sqrt{\frac{2}{k}} \phi +
  \sqrt{\frac{k+2}{k}} T \right)}
\ee
which satisfy an $SL(2,R)$ algebra at level $k+2$
:\footnote{If we wish, we
can add a super-partner for the boson $T$,
 and
view the linear combination of $\d \phi$ and $\d T$ orthogonal
to
$I^3$ as bosonized complex fermion:
\be
i \sqrt{\frac{k+2}{k}} \d \phi + \sqrt{\frac{2}{k}} \d T  =
:\psi^+ \psi^-:.
\ee
The three fermions then complete an $N=1$ supersymmetric \slr$_k$ theory,
containing a purely bosonic \slr$_{k+2}$.}
\eqn{opeads}{\eqalign{
j^3(z) j^3(w) \sim & \, -{k+2 \over 2(z-w)^2} , \cr
j^3(z) j^{\pm}(w) \sim & \, \pm { j^{\pm}(w) \over z-w}, \cr
j^+(z) j^-(w) \sim & \, {k+2 \over (z-w)^2} - {2 j^3(w) \over z-w}.
}}

To make use of the purely bosonic
 current algebra when solving the theory, we will need
to introduce primary fields for the $N=2$ algebra combined with the $U(1)$
that transform as standard representations of the $SL(2,R)$ current
algebra. This can be achieved as follows. When we denote $Z_{j;r,\bar{r}}$
a primary of the $N=2$ algebra with conformal weight $\Delta_{j,r}
= -j(j-1)/k + m^2/k$
 and $U(1)$ R-charge
$m=j+r$,
we can introduce the new primaries $\hat{Z}_{j,r,\bar{r}}$:
\be
\hat{Z}_{j,r,\bar{r}}= Z_{j,r,\bar{r}}e^{\alpha_{j,r}T_L + \alpha_{j,\bar{r}}T_R},
\ee
where we choose $\alpha_{j,r}$ such that
\be
\Delta(\hat{Z}_{j,r,\bar{r}}) =-{j(j-1) \over k} =  \Delta_{j,r} - {\alpha_{j,r}^2 \over 2}
\ee
(i.e. the momentum of the boson
 $T$ is coupled to the $N=2$ $U(1)$ R-charge).
The new primaries can be checked to transform in a standard fashion as operators of
$SL(2,R)$, in a basis labeled by the spectrum of an elliptic generator. To
go to the hyperbolic basis we can Fourier transform:
\be
\Phi_j (z,\bar{z},x,\bar{x})= \sum_{m,\bar{m}} \hat{Z}_{j,r,\bar{r}} \,\, x^{j-1+m} \bar{x}^{j-1+\bar{m}}.
\ee
We can now compute correlators of
these fields, which are linear combinations of primaries of the $N=2$ algebra
"dressed" with an orthogonal and free $U(1)$, using a full chiral $SL(2,R)$.
Thus, if  the solution of the $H_3^{+}$ theory is purely based on
symmetries, this holds for  any $N=2$ theory with $c>3$. \footnote{We
assume that the $N=2$ theory does not split into subtheories with $c<3$.
This assumes the irreducibility of the representations space of the $N=2,c>3$
algebra. Otherwise, a similar $SU(2)$ algebra would be a
 more appropriate tool to solve the theory.}

We note also that the construction above shows that the "$x$-variables" basis
will remain an efficient formalism for computation as long as $N=2$
supersymmetry is preserved. This complements the remarks in
\cite{Aharony:2004xn} where it was suggested that these variables should be interpreted as
parameterizing a new worldsheet, at any point in the moduli space.

\section{Conventions}
We take $q=e^{i 2 \pi \tau}, y=e^{i 2 \pi \nu}$, and we define:
\be
\eta(\tau) &=& q^{1/24} \prod_{n=1}^{\infty} (1-q^{n}) \\
\vartheta_2(\tau,\nu ) &=&
q^{\frac18}y^{\frac12} \prod_{n=1}^{\infty}(1-q^n)(1+q^{n}y)(1+q^{n-1}y^{-1}) =
\sum_{n \in \mathbb{Z}+\frac12} q^{n^2 \over 2}y^n
\label{te2} \\
\vartheta_3(\tau,\nu ) &=&
\prod_{n=1}^{\infty}(1-q^n)(1+q^{n-\frac12}y)(1+q^{n-\frac12}y^{-1}) =
\sum_{n \in \mathbb{Z}} q^{n^2 \over 2}y^n
\label{te3} \\
\vartheta_4(\tau,\nu ) &=&
\prod_{n=1}^{\infty}(1-q^n)(1-q^{n-\frac12}y)(1-q^{n-\frac12}y^{-1}) =
\sum_{n \in \mathbb{Z}} (-1)^n q^{n^2 \over 2}y^n.
\label{te4}
\ee
A shift in the second argument of the third theta-function  can be compensated
for as follows $(w \in \mathbb{Z})$:
\be
\vartheta_3(\tau,\eta+w\tau)=q^{-\frac{w^2}{2}}y^{-w}\vartheta_3(\tau,\eta).
\label{flowtheta}
\ee

%

\begin{thebibliography}{99}


\bibitem{Girardello:1990sh}
L.~Girardello, A.~Pasquinucci and M.~Porrati, ``N=2
Morse-Liouville Theory And Nonminimal Superconformal Theories,''
Nucl.\ Phys.\  {\bf B352}, 769 (1991).

\bibitem{Kutasov:1990ua}
D.~Kutasov and N.~Seiberg, ``Noncritical Superstrings,'' Phys.\
Lett.\  {\bf B251}, 67 (1990).

\bibitem{Kazama:1988qp}
Y.~Kazama and H.~Suzuki, ``New N=2 Superconformal Field Theories
And Superstring Compactification,'' Nucl.\ Phys.\  {\bf B321}, 232
(1989).

\bibitem{FZZ}
V. Fateev, A. Zamolodchikov, A. Zamolodchikov, unpublished notes.


\bibitem{Giveon:1999px}
A.~Giveon and D.~Kutasov, ``Little string theory in a double
scaling limit,'' JHEP {\bf 9910}, 034 (1999)
[arXiv: hep-th/9909110].

\bibitem{Hori:2001ax}
K.~Hori and A.~Kapustin, ``Duality of the fermionic 2d black hole
and N = 2 Liouville theory as  mirror symmetry,'' JHEP {\bf 0108},
045 (2001) [arXiv: hep-th/0104202].

\bibitem{Tong:2003ik}
D.~Tong, ``Mirror mirror on the wall: On two-dimensional black
holes and Liouville theory,'' JHEP {\bf 0304}, 031 (2003)
[arXiv: hep-th/0303151].

\bibitem{Israel:2004ir}
D.~Israel, C.~Kounnas, A.~Pakman and J.~Troost, ``The partition
function of the supersymmetric two-dimensional black hole and
little string theory,'' arXiv: hep-th/0403237.


\bibitem{Seiberg:1997zk}
N.~Seiberg, ``New theories in six dimensions and matrix
description of M-theory on  T**5 and T**5/Z(2),'' Phys.\ Lett.\
{\bf B408}, 98 (1997) [arXiv:hep-th/9705221];
M.~Berkooz, M.~Rozali and N.~Seiberg,
``On transverse fivebranes in M(atrix) theory on T**5,''
Phys.\ Lett.\  {\bf B408}, 105 (1997)
[arXiv: hep-th/9704089].


\bibitem{Eguchi:2003ik}
T.~Eguchi and Y.~Sugawara, ``Modular bootstrap for boundary N = 2
Liouville theory,'' JHEP {\bf 0401}, 025 (2004)
[arXiv: hep-th/0311141].



\bibitem{McGreevy:2003dn}
J.~McGreevy, S.~Murthy and H.~Verlinde, ``Two-dimensional
superstrings and the supersymmetric matrix model,'' JHEP {\bf
0404}, 015 (2004) [arXiv: hep-th/0308105].

\bibitem{Ahn:2003tt}
C.~Ahn, M.~Stanishkov and M.~Yamamoto, ``One-point functions of N
= 2 super-Liouville theory with boundary,'' Nucl.\ Phys.\
{\bf B683}, 177 (2004) [arXiv: hep-th/0311169].

\bibitem{Giveon:2003wn}
A.~Giveon, A.~Konechny, A.~Pakman and A.~Sever, ``Type 0 strings
in a 2-d black hole,'' JHEP {\bf 0310}, 025 (2003)
[arXiv:hep-th/0309056].


\bibitem{Douglas:2003up}
M.~R.~Douglas, I.~R.~Klebanov, D.~Kutasov, J.~Maldacena, E.~Martinec and N.~Seiberg,
``A new hat for the c = 1 matrix model,''
arXiv:hep-th/0307195.

\bibitem{Kazakov:2000pm}
V.~Kazakov, I.~K.~Kostov and D.~Kutasov, ``A matrix model for the
two-dimensional black hole,'' Nucl.\ Phys.\  {\bf B622}, 141
(2002) [arXiv: hep-th/0101011].



\bibitem{Elitzur:cb}
S.~Elitzur, A.~Forge and E.~Rabinovici, ``Some Global Aspects Of
String Compactifications,'' Nucl.\ Phys.\  {\bf B359}, 581 (1991).

\bibitem{Mandal:1991tz}
G.~Mandal, A.~M.~Sengupta and S.~R.~Wadia, ``Classical solutions
of two-dimensional string theory,'' Mod.\ Phys.\ Lett.\ {\bf A6},
1685 (1991).

\bibitem{Witten:1991yr}
E.~Witten, ``On string theory and black holes,''
Phys.\ Rev.\ {\bf D44}, 314 (1991).

\bibitem{Giveon:1991sy}
A.~Giveon, ``Target space duality and stringy black holes,'' Mod.\
Phys.\ Lett.\ {\bf A6}, 2843 (1991).


\bibitem{Dijkgraaf:1991ba}
R.~Dijkgraaf, H.~Verlinde and E.~Verlinde, ``String propagation in
a black hole geometry,'' Nucl.\ Phys.\ {\bf B371}, 269 (1992).


\bibitem{Henningson:1991jc}
M.~Henningson, S.~Hwang, P.~Roberts and B.~Sundborg, ``Modular
invariance of SU(1,1) strings,'' Phys.\ Lett.\ {\bf B267}, 350
(1991).



\bibitem{Maldacena:2000hw}
J.~M.~Maldacena and H.~Ooguri, ``Strings in AdS(3) and SL(2,R) WZW
model. I,'' J.\ Math.\ Phys.\  {\bf 42}, 2929 (2001)
[arXiv: hep-th/0001053].

\bibitem{Dixon:1989cg}
L.~J.~Dixon, M.~E.~Peskin and J.~Lykken, ``N=2 Superconformal
Symmetry And SO(2,1) Current Algebra,'' Nucl.\ Phys.\ {\bf B325},
329 (1989).


\bibitem{Hanany:2002ev}
A.~Hanany, N.~Prezas and J.~Troost, ``The partition function of
the two-dimensional black hole conformal  field theory,'' JHEP
{\bf 0204}, 014 (2002) [arXiv: hep-th/0202129].


\bibitem{Israel:2003ry}
D.~Israel, C.~Kounnas and M.~P.~Petropoulos,
``Superstrings on NS5 backgrounds, deformed AdS(3) and holography,''
JHEP {\bf 0310}, 028 (2003)
[arXiv:hep-th/0306053].

\bibitem{Eguchi:2004yi}
T.~Eguchi and Y.~Sugawara, ``SL(2,R)/U(1) supercoset and elliptic
genera of non-compact Calabi-Yau manifolds,''
JHEP {\bf 0405} (2004) 014
[arXiv: hep-th/0403193].

\bibitem{Becker:1993at}
K.~Becker and M.~Becker, ``Interactions in the SL(2,IR) / U(1)
black hole background,'' Nucl.\ Phys.\  {\bf B418}, 206 (1994)
[arXiv: hep-th/9310046].


\bibitem{Dotsenko:ui}
V.~S.~Dotsenko, ``The Free Field Representation Of The SU(2)
Conformal Field Theory,'' Nucl.\ Phys.\  {\bf B338}, 747 (1990);
V.~S.~Dotsenko, ``Solving The SU(2) Conformal Field Theory With
The Wakimoto Free Field Representation,'' Nucl.\ Phys.\
{\bf B358}, 547 (1991).


\bibitem{Giribet:2001ft}
G.~Giribet and C.~Nunez, ``Correlators in AdS(3) string theory,''
JHEP {\bf 0106}, 010 (2001) [arXiv: hep-th/0105200].

\bibitem{Hofman:2004ny}
D.~M.~Hofman and C.~A.~Nunez, ``Free field realization of
superstring theory on AdS3,'' arXiv: hep-th/0404214.


\bibitem{Baseilhac:1998eq}
P.~Baseilhac and V.~A.~Fateev, ``Expectation values of local
fields for a two-parameter family of  integrable models and
related perturbed conformal field theories,'' Nucl.\ Phys.\
{\bf B532}, 567 (1998) [arXiv: hep-th/9906010].

\bibitem{Fukuda:2001jd}
T.~Fukuda and K.~Hosomichi, ``Three-point functions in
sine-Liouville theory,'' JHEP {\bf 0109}, 003 (2001)
[arXiv: hep-th/0105217].

\bibitem{Giribet:2004zd}
G.~Giribet and D.~Lopez-Fogliani, ``Remarks on free field
realization of SL(2,R)/U(1) x U(1) WZNW model,''
arXiv: hep-th/0404231.

\bibitem{Maldacena:2001km}
J.~M.~Maldacena and H.~Ooguri, ``Strings in AdS(3) and the SL(2,R)
WZW model. III: Correlation  functions,'' Phys.\ Rev.\  {\bf D65},
106006 (2002) [arXiv: hep-th/0111180].

\bibitem{Giveon:2001up}
A.~Giveon and D.~Kutasov, ``Notes on AdS(3),'' Nucl.\ Phys.\ B
{\bf 621}, 303 (2002) [arXiv:hep-th/0106004].

\bibitem{Hori:2000kt}
K.~Hori and C.~Vafa,
``Mirror symmetry,''
arXiv: hep-th/0002222.



\bibitem{Ahn:2002sx}
C.~Ahn, C.~Kim, C.~Rim and M.~Stanishkov, ``Duality in N = 2
super-Liouville theory,'' arXiv:hep-th/0210208.

\bibitem{Nakayama:2004vk}
Y.~Nakayama, ``Liouville field theory: A decade after the
revolution,'' arXiv: hep-th/0402009.

\bibitem{Green:1987qu}
M.~B.~Green and N.~Seiberg,
``Contact Interactions In Superstring Theory,''
Nucl.\ Phys.\ {\bf B299}, 559 (1988).

\bibitem{Aharony:2004xn}
O.~Aharony, A.~Giveon and D.~Kutasov, ``LSZ in LST,''
arXiv: hep-th/0404016.

\bibitem{Goulian:1990qr}
M.~Goulian and M.~Li, ``Correlation Functions In Liouville
Theory,'' Phys.\ Rev.\ Lett.\  {\bf 66}, 2051 (1991);
 P.~Di Francesco and D.~Kutasov,
``World sheet and space-time physics in two-dimensional
(Super)string theory,'' Nucl.\ Phys.\  {\bf B375}, 119 (1992)
[arXiv:hep-th/9109005].

\bibitem{Teschner:2001rv}
J.~Teschner, ``Liouville theory revisited,'' Class.\ Quant.\
Grav.\  {\bf 18}, R153 (2001) [arXiv: hep-th/0104158].

\bibitem{Ribault:2003ss}
S.~Ribault and V.~Schomerus, ``Branes in the 2-D black hole,''
JHEP {\bf 0402}, 019 (2004) [arXiv: hep-th/0310024].



\bibitem{Fukuda:2002bv}
T.~Fukuda and K.~Hosomichi,
``Super Liouville theory with boundary,''
Nucl.\ Phys.\  {\bf B635}, 215 (2002)
[arXiv: hep-th/0202032].

\bibitem{Zamolodchikov:2001ah}
A.~B.~Zamolodchikov and A.~B.~Zamolodchikov, ``Liouville field
theory on a pseudosphere,'' arXiv: hep-th/0101152.

\bibitem{Fateev:2000ik}
V.~Fateev, A.~B.~Zamolodchikov and A.~B.~Zamolodchikov, ``Boundary
Liouville field theory. I: Boundary state and boundary  two-point
function,'' arXiv: hep-th/0001012; \,
J.~Teschner, ``Remarks on Liouville theory with boundary,''
arXiv: hep-th/0009138.


\bibitem{Teschner:1995yf}
J.~Teschner, ``On the Liouville three point function,'' Phys.\
Lett.\  {\bf B363}, 65 (1995) [arXiv: hep-th/9507109]; \,
J.~Teschner, ``On structure constants and fusion rules in the
SL(2,C)/SU(2) WZNW  model,'' Nucl.\ Phys.\ {\bf B546}, 390 (1999)
[arXiv: hep-th/9712256].

\bibitem{Lee:2001gh}
P.~Lee, H.~Ooguri and J.~w.~Park,
``Boundary states for AdS(2) branes in AdS(3),''
Nucl.\ Phys.\ B {\bf 632}, 283 (2002)
[arXiv:hep-th/0112188].


\bibitem{Ponsot:2001gt}
B.~Ponsot, V.~Schomerus and J.~Teschner, ``Branes in the Euclidean
AdS(3),'' JHEP {\bf 0202}, 016 (2002) [arXiv: hep-th/0112198].

\bibitem{Ooguri:1996ck}
H.~Ooguri, Y.~Oz and Z.~Yin, ``D-branes on Calabi-Yau spaces and
their mirrors,'' Nucl.\ Phys.\ B {\bf 477} (1996) 407
[arXiv: hep-th/9606112].

\bibitem{Bachas:2000fr}
C.~Bachas and M.~Petropoulos, ``Anti-de-Sitter D-branes,'' JHEP
{\bf 0102} (2001) 025 [arXiv: hep-th/0012234].

\bibitem{Rhedin:1995um}
H.~Rhedin, ``Gauged supersymmetric WZNW model using the BRST
approach,'' Phys.\ Lett.\  {\bf B373} (1996) 76
[arXiv: hep-th/9511143].



\bibitem{Fotopoulos:2003vc}
A.~Fotopoulos,
``Semiclassical description of D-branes in SL(2)/U(1) gauged WZW model,''
Class.\ Quant.\ Grav.\  {\bf 20}, S465 (2003)
[arXiv: hep-th/0304015].

\bibitem{Matsuo:1986vc}
Y.~Matsuo and S.~Yahikozawa, ``Superconformal Field Theory With
Modular Invariance On A Torus,'' Phys.\ Lett.\  {\bf B178}, 211
(1986).

\bibitem{Giveon:2001uq}
A.~Giveon, D.~Kutasov and A.~Schwimmer, ``Comments on D-branes in
AdS(3),'' Nucl.\ Phys.\  {\bf B615}, 133 (2001)
[arXiv: hep-th/0106005].


\bibitem{Miki:1989ri}
K.~Miki, ``The Representation Theory Of The SO(3) Invariant
Superconformal Algebra,'' Int.\ J.\ Mod.\ Phys.\ {\bf A5}, 1293
(1990).


\bibitem{Israel:2004xj}
D.~Israel, A.~Pakman and J.~Troost, ``Extended SL(2,R)/U(1)
characters, or modular properties of a simple non-rational
conformal field theory,''
JHEP {\bf 0404} (2004) 045
[arXiv: hep-th/0402085].

\bibitem{Recknagel:1997sb}
A.~Recknagel and V.~Schomerus,
``D-branes in Gepner models,''
Nucl.\ Phys.\ B {\bf 531}, 185 (1998)
[arXiv:hep-th/9712186].

\bibitem{Bakas:1991fs}
I.~Bakas and E.~Kiritsis, ``Beyond the large N limit: Nonlinear
W(infinity) as symmetry of the SL(2,R) / U(1) coset model,'' Int.\
J.\ Mod.\ Phys.\ {\bf A7}, 55 (1992) [arXiv: hep-th/9109029].


\bibitem{Ribault:2002ti}
S.~Ribault, ``Two AdS(2) branes in the Euclidean AdS(3),'' JHEP
{\bf 0305}, 003 (2003) [arXiv:hep-th/0210248].

\bibitem{Martinec:2003ka}
E.~J.~Martinec,
``The annular report on non-critical string theory,''
arXiv:hep-th/0305148.

\bibitem{Seiberg:2003nm}
N.~Seiberg and D.~Shih,
``Branes, rings and matrix models in minimal (super)string theory,''
JHEP {\bf 0402}, 021 (2004)
[arXiv:hep-th/0312170].


\bibitem{Elitzur:zh}
S.~Elitzur, A.~Giveon, D.~Kutasov, E.~Rabinovici and G.~Sarkisian,
``D-Branes In The Background Of Ns Fivebranes,'' Int.\ J.\ Mod.\
Phys.\  {\bf A16} (2001) 880.

\bibitem{Pelc:2000kb}
O.~Pelc, ``On the quantization constraints for a D3 brane in the
geometry of NS5 branes,'' JHEP {\bf 0008}, 030 (2000)
[arXiv: hep-th/0007100].

\bibitem{Ribault:2003sg}
S.~Ribault, ``D3-branes in NS5-branes backgrounds,'' JHEP {\bf
0302}, 044 (2003) [arXiv: hep-th/0301092].


\bibitem{LSTbranes}
D.~Israel, A.~Pakman and J.~Troost,
``D-branes in Little String Theory'', to appear.


\bibitem{Dobrev:wd}
V.~K.~Dobrev, ``Structure Of Verma Modules And Characters Of
Irreducible Highest Weight Modules Over N=2 Superconformal
Algebras,'' in "CLAUSTHAL 1986, PROCEEDINGS, DIFFERENTIAL
GEOMETRIC METHODS IN THEORETICAL PHYSICS", 289-307.

\bibitem{Dobrev:1986hq}
V.~K.~Dobrev, ``Characters Of The Unitarizable Highest Weight
Modules Over The N=2 Superconformal Algebras,'' Phys.\ Lett.\ {\bf
B186}, 43 (1987).

\bibitem{Kiritsis:1986rv}
E.~Kiritsis, ``Character Formulae And The Structure Of The
Representations Of The N=1, N=2 Superconformal Algebras,'' Int.\
J.\ Mod.\ Phys.\ {\bf A3}, 1871 (1988).


\bibitem{Goddard:vk}
P.~Goddard, A.~Kent and D.~I.~Olive, ``Virasoro Algebras And Coset
Space Models,'' Phys.\ Lett.\  {\bf B152}, 88 (1985);
P.~Goddard, A.~Kent and D.~I.~Olive, ``Unitary Representations Of
The Virasoro And Supervirasoro Algebras,'' Commun.\ Math.\ Phys.\
{\bf 103}, 105 (1986);
Z.~a.~Qiu,
``Modular Invariant Partition Functions For N=2 Superconformal Field
Theories,'' Phys.\ Lett.\ B {\bf 198}, 497 (1987).



\bibitem{Schwimmer:mf}
A.~Schwimmer and N.~Seiberg, ``Comments On The N=2, N=3, N=4
Superconformal Algebras In Two-Dimensions,'' Phys.\ Lett.\
{\bf B184}, 191 (1987).

\bibitem{Boucher:1986bh}
W.~Boucher, D.~Friedan and A.~Kent, ``Determinant Formulae And
Unitarity For The N=2 Superconformal Algebras In Two-Dimensions Or
Exact Results On String Compactification,'' Phys.\ Lett.\
{\bf B172}, 316 (1986).


\bibitem{Pakman:2003cu}
A.~Pakman, ``Unitarity of supersymmetric SL(2,R)/U(1) and no-ghost
theorem for fermionic strings in AdS(3) x N,'' JHEP {\bf 0301},
077 (2003) [arXiv: hep-th/0301110].


\bibitem{Pakman:2003kh}
A.~Pakman, ``BRST quantization of string theory in AdS(3),'' JHEP
{\bf 0306}, 053 (2003) [arXiv: hep-th/0304230].

\bibitem{Douglas:1999hq}
M.~R.~Douglas and B.~Fiol,
``D-branes and discrete torsion. II,''
arXiv: hep-th/9903031.





\end{thebibliography}
\end{document}